\documentclass[11pt,a4paper]{article}
\pdfoutput=1
\usepackage{jheppub}
\usepackage{slashed}

\makeatletter
\def\@fpheader{\relax}
\makeatother

\usepackage{ulem}

\usepackage[czech,english]{babel}
\usepackage{graphicx}
\usepackage{amsmath,amsfonts,amssymb}
\usepackage{url}
\usepackage{mathabx}
\DeclareMathOperator{\MyProd}{\scalebox{1.4}{$\mathrm{I\kern-0.2ex I}$}}

\preprint{LCTP-19-22}

\title{Microscopic Entropy of Rotating Electrically Charged AdS$_4$ Black Holes from Field Theory Localization}

\author[a]{Jun Nian}

\emailAdd{nian@umich.edu}

\affiliation[a]{Leinweber Center for Theoretical Physics, University of Michigan, Ann Arbor, MI 48109, U.S.A.}

\author[a, b]{and Leopoldo A. Pando Zayas}

\emailAdd{lpandoz@umich.edu}

\affiliation[b]{The Abdus Salam International Centre for Theoretical Physics, 34014 Trieste, Italy}

\abstract{We employ  supersymmetric localization to determine the exact partition function of 3d $\mathcal{N}=2$ gauge theories on a background given by a round $S^2$ fibered over a circle and certain complexified background fields. The Coulomb branch localization locus includes monopole  configurations, and the partition function reduces to a matrix model. We consider the partition function of the ABJM theory on this background as an explicit case. We verify that the large-$N$ limit of the ABJM theory  partition function produces, in the Cardy limit,  the entropy function of the dual rotating, electrically charged asymptotically AdS$_4$ supersymmetric black holes and thus provides a microscopic explanation for the Bekenstein-Hawking entropy.}


\arxivnumber{}

\newcommand{\bea}{\begin{eqnarray}}
\newcommand{\eea}{\end{eqnarray}}

\newcommand{\be}{\begin{equation}}
\newcommand{\ee}{\end{equation}}

\begin{document}

\maketitle

%%%%%%%%%%%%%%%%%%%%%%%%%%%%%%%%%%%%%%%%%%%
%%%%%%%%%%%%%%%%%%%%%%%%%%%%%%%%%%%%%%%%%%%
\section{Introduction}\label{sec:introduction}
%%%%%%%%%%%%%%%%%%%%%%%%%%%%%%%%%%%%%%%%%%%
%%%%%%%%%%%%%%%%%%%%%%%%%%%%%%%%%%%%%%%%%%%

String theory, in the ground-breaking works of Strominger and Vafa \cite{Strominger:1996sh},  has demonstrated its viability as a theory of quantum gravity by providing a microscopic interpretation for the macroscopic entropy of certain black holes. A similar microscopic explanation for asymptotically AdS black holes has long been an important open problem. The AdS/CFT correspondence \cite{Maldacena:1997re, Witten:1998qj, Gubser:1998bc} provides us with some insights into black hole puzzles by identifying a gravity theory in asymptotically AdS spacetimes with a field theory on the boundary. It, therefore, implicitly answers questions such as the information paradox via a unitary field theory; we would like to know the answers explicitly.

The breakthrough occurred  first for  magnetically charged asymptotically AdS$_4$ black holes \cite{Benini:2015eyy}. Benini, Hristov and Zaffaroni studied the topologically twisted index of the dual ABJM theory on the boundary, and found that at the leading order it matches exactly the black hole entropy of AdS$_4$ magnetically charged STU  black holes after a Legendre transformation. Since then many more cases and generalizations have been considered \cite{Benini:2016rke, Cabo-Bizet:2017jsl, Benini:2017oxt,Hosseini:2017fjo,Azzurli:2017kxo,Hosseini:2016cyf, Toldo:2017qsh,Gang:2018hjd,Hosseini:2018uzp,Crichigno:2018adf,Fluder:2019szh,Hong:2018viz} including incursions into sub-leading corrections \cite{Liu:2017vll,Jeon:2017aif,Liu:2017vbl,Hristov:2018lod, Liu:2018bac,Gang:2019uay, PandoZayas:2019hdb,Hristov:2019xku}.

For the asymptotically AdS$_5$ case, using the superconformal index of $\mathcal{N}=4$ super-Yang-Mills (SYM) theory on the boundary, some attempts of reproducing the AdS$_5$ black hole entropy via the AdS/CFT correspondence have been made \cite{Kinney:2005ej}, but the results did not quite match the ones from the gravity side.

Recently, it was first found, on the gravity side, that the entropies of $1/16$-BPS AdS$_5$ electrically charged black holes \cite{Gutowski:2004ez, Gutowski:2004yv, Chong:2005da, Chong:2005hr, Kunduri:2006ek, Kim:2006he} can be obtained by extremizing an entropy function \cite{Hosseini:2017mds}, which has the same functional
form as the supersymmetric Casimir energy of $\mathcal{N}=4$ SYM first studied in \cite{Bobev:2015kza}. Several groups have now independently reproduced the entropy function by studying the partition function or the superconformal index of $\mathcal{N}=4$ SYM \cite{Cabo-Bizet:2018ehj, Choi:2018hmj, Benini:2018ywd} by slightly reinterpreting the original work in the superconformal index \cite{Kinney:2005ej}. Further extensions for the general growth of the  $\mathcal{N}=1$ superconformal index \cite{Honda:2019cio, ArabiArdehali:2019tdm, Kim:2019yrz, Cabo-Bizet:2019osg, Amariti:2019mgp} including via the Bethe-Ansatz approach \cite{Lezcano:2019pae, Lanir:2019abx} have been achieved. After this important progress, the generalizations to other dimensions were also considered. For instance, the BPS AdS$_6$ and AdS$_4$ black hole entropy functions have been studied on the gravity side in \cite{Choi:2018fdc, Cassani:2019mms}, and computed from the microscopic theories using the superconformal index \cite{Choi:2019miv, Choi:2019zpz, Bobev:2019zmz,Benini:2019dyp}. Similarly, the BPS AdS$_7$ black hole entropy obtained from gravity \cite{Hosseini:2018dob} was also reproduced recently using dual field theory partition function \cite{Kantor:2019lfo} and superconformal index \cite{Nahmgoong:2019hko}. By turning on temperature or breaking the BPS constraint, the near-BPS AdS$_5$ black hole entropy was also computed both on the gravity side and from the microscopic field-theoretic side \cite{Larsen:2019oll}.

Although a lot of progress has been made towards the understanding of AdS black hole entropies in various dimensions, most of the recent works are based on the study of superconformal indices, which  are essentially computations done in the free field theory  limit. Hence, the results obtained in this way capture only perturbative information, and are probably only reliable at the leading order. In order to go beyond the perturbative results at the leading order, we need some exact non-perturbative approaches, and supersymmetric localization is  such a tool. Among the recent progress, it was first introduced in \cite{Cabo-Bizet:2018ehj} that the physical partition function of $\mathcal{N}=4$ SYM without a topological twist can be computed on certain complex backgrounds using supersymmetric localization. The result was used to obtain the BPS AdS$_5$ black hole entropy function. In this paper, we extend  this idea to the BPS AdS$_4$ black hole case by computing the exact partition function of 3d ABJM theory on an appropriate curved background geometry that includes complexified background fields. The result matches the BPS AdS$_4$ entropy function from the gravity side \cite{Choi:2018fdc}. Similarly and alternatively to  \cite{Choi:2019zpz} our computation provides a microscopic foundation for the Bekenstein-Hawking entropy of such black holes. 

The first technical difficulty is the construction  of the field theory on a curved  background including complexified  background fields  on which the field theory can be defined in a supersymmetric way. The systematic approach to this problem was formulated in the work of  Festuccia and Seiberg \cite{Festuccia:2011ws} who starting from a supergravity theory took a rigid limit leading to supersymmetric field theories on curved spaces.  This general approach was explicitly applied to 3-dimensional curved spaces \cite{Closset:2012ru}, and some specific complex backgrounds were constructed in \cite{Nian:2013qwa} in order to produce the partition functions of 3d $\mathcal{N}=2$ Chern-Simons-matter theories on squashed $S^3$, elucidating a previous  puzzle regarding two theories with the same supersymmetry but with different partition functions on the same curved space as obtained in \cite{Hama:2011ea, Imamura:2011wg}.

In this paper, we apply the same technique of \cite{Nian:2013qwa} to the boundary of rotating  electrically charged supersymmetric  AdS$_4$ black holes, which is a round $S^2$ fibered over  a circle, $S^1$, and described by the metric 
\be
  ds^2 = d\tau^2 + L^2 \Big[d\theta^2 + \textrm{sin}^2 \theta \, (d\varphi - i \Omega d\tau)^2\Big]\, .
\ee
By appropriately turning on some complex background fields, we find that there exist Killing spinors with anti-periodic boundary conditions along the circle $S^1$ which is a requirement for supersymmetric localization.  The background is characterized by the chemical potential $\Omega$ for the angular momentum and the electric potential $\Phi$, both of which are in general complex-valued, and there is a constraint between $\Omega$ and $\Phi$ in order to preserve some supercharges. Using some supercharges on this background, we apply supersymmetric localization techniques to compute the partition function of 3d $\mathcal{N}=2$ Chern-Simons-matter theories. To address the ABJM theory we lift the degeneracy from $L \Phi$ to $\Delta_I / 2$ ($I = 1, \cdots, 4$) by turning on chemical potentials for the flavor symmetry. We study the large-$N$ limit of the free energy of the ABJM theory and show that if one additionally restricts to a Cardy-like limit  ($|\omega| \equiv |L \Omega| \ll 1$), the large-$N$ free energy  has the expression:
\be
  F \simeq \frac{2 \sqrt{2}\, i\, k^{\frac{1}{2}} N^{\frac{3}{2}}}{3} \frac{\sqrt{\Delta_1 \Delta_2 \Delta_3 \Delta_4}}{\omega}\, ,
\ee
which precisely  reproduces the entropy function of the rotating electrically charged BPS AdS$_4$ black holes \cite{Choi:2018fdc, Cassani:2019mms}. Our microscopic derivation is based on an approach quite different from \cite{Choi:2019zpz} whose starting point is the superconformal index.

The manuscript is organized as follows. In section~\ref{sec:BgdFields} we describe the general setup which includes a curved three-dimensional metric as well as complex background fields, in a manner that admits certain amount of supersymmetry and Killing spinors with anti-periodic boundary condition along $S^1$. In that section we also briefly review the supersymmetric gauge theories that can be defined on the constructed backgrounds. In section~\ref{sec:Localization} the partition function of 3d $\mathcal{N}=2$ gauge theories are computed using supersymmetric localization. Various  details of the general computation are presented and we discuss the ABJM theory as an explicit case. In section~\ref{sec:LargeN} we  compute the free energy in the  large-$N$ and further take a Cardy-like limit. In section~\ref{sec:BHEntropyFct} we demonstrate that the free energy of the ABJM theory in that large-$N$ and a Cardy-like limit reproduces the Bekenstein-Hawking entropy of rotating electrically charged supersymmetric  AdS$_4$ black hole, thus providing a counting of microstates via the AdS/CFT correspondence. A discussion, including prospects for future research, is presented in section~\ref{sec:Discussion}. Some conventions of spinors are summarized in appendix~\ref{app:Convention}, useful identities of special functions are listed in appendix~\ref{app:SpecialFct}.

%%%%%%%%%%%%%%%%%%%%%%%%%%%%%%%%%%%%%%%%%%%
%%%%%%%%%%%%%%%%%%%%%%%%%%%%%%%%%%%%%%%%%%%
\section{3d $\mathcal{N}=2$ SUSY Theories on Curved Spaces}\label{sec:BgdFields}
%%%%%%%%%%%%%%%%%%%%%%%%%%%%%%%%%%%%%%%%%%%
%%%%%%%%%%%%%%%%%%%%%%%%%%%%%%%%%%%%%%%%%%%

\subsection{Supersymmetry with Background Fields}

An important requisite for supersymmetric localization is to establish supersymmetry on the given curved background. The term curved background not only refers to the metric but to the set of auxiliary background fields involved.

The case we are mainly interested in has the following metric:
\be\label{eq:metric-1}
  ds^2 = d\tau^2 + L^2 \Big[d\theta^2 + \textrm{sin}^2 \theta \, (d\varphi - i \Omega d\tau)^2\Big]\, ,
\ee
where $\Omega$ is a purely imaginary constant,  such that \eqref{eq:metric-1} is a real metric,  and $\tau$ is a coordinate with period $L$,  that is,
\be
  \tau \sim \tau + L\, .
\ee
We choose the period of $S^1$ and the radius of $S^2$ to be both $L$.

We can rewrite the metric as
\be\label{eq:metric-2}
  ds^2 = f^2 (d\chi + a dz + \bar{a} d\bar{z})^2 + c^2 dz d\bar{z}\, ,
\ee
where $z$ and $\bar{z}$ denote the complex coordinates on the sphere $S^2$. More precisely,
\begin{align}
  z & = \textrm{tan} \frac{\theta}{2}\, e^{i (\varphi - i \Omega \tau)}\, ,\\
  \chi & = \tau\, ,
\end{align}
and the factors in the metric \eqref{eq:metric-2} are chosen to be
\begin{align}
  f & = 1\, ,\\
  a & = 0\, ,\\
  c & = L\, \left(1 + \textrm{cos}\, \theta \right)\, .\label{eq:def-c}
\end{align}
For later convenience, we also define another set of coordinates:
\begin{align}
  \widetilde{\tau} & \equiv \tau\, ,\\
  \widetilde{\theta} & \equiv \theta\, ,\\
  \widetilde{\varphi} & \equiv \varphi - i \Omega \tau\, .
\end{align}
In these new coordinates the metric \eqref{eq:metric-1} becomes
\be\label{eq:metric-3}
  ds^2 = d\widetilde{\tau}^2 + L^2 \Big[d\widetilde{\theta}^2 + \textrm{sin}^2 \theta \, d\widetilde{\varphi}^2\Big]\, ,
\ee
showing its local equivalence to the standard metric on $S^1 \times S^2$. Globally,  the difference is encoded in the identifications of the coordinates.

One systematic approach to establishing supersymmetry for field theories in curved backgrounds  is to start with  a supergravity theory and then consider its rigid limit  \cite{Festuccia:2011ws}. Following closely the discussion in Ref.~\cite{Closset:2012ru} which is rooted in the minimal gauged supergravity in three dimensions, one can construct two conserved supercharges satisfying the following Killing spinor equations stemming from the gravitino variations:
\begin{align}
  (\nabla_\mu - i A_\mu) \zeta & = - \frac{1}{2} H \gamma_\mu \zeta - i V_\mu \zeta - \frac{1}{2} \epsilon_{\mu\nu\rho} V^\nu \gamma^\rho \zeta\, ,\label{eq:KillingSpEq-1}\\
  (\nabla_\mu + i A_\mu) \widetilde{\zeta} & = - \frac{1}{2} H \gamma_\mu \widetilde{\zeta} + i V_\mu \widetilde{\zeta} + \frac{1}{2} \epsilon_{\mu\nu\rho} V^\nu \gamma^\rho \widetilde{\zeta}\, ,\label{eq:KillingSpEq-2}
\end{align}
where
\be\label{eq:KillingSp-1}
  \zeta_\alpha = \sqrt{s} \left( \begin{array}{c}
  1 \\ 0
  \end{array} \right)\, ,\quad
  \zeta^\alpha = \sqrt{s} \left( \begin{array}{c}
  0 \\ -1
  \end{array} \right)\, .
\ee
\be\label{eq:KillingSp-2}
  \widetilde{\zeta}_\alpha = \frac{1}{\sqrt{s}} \left( \begin{array}{c}
  0 \\ 1
  \end{array} \right)\, ,\quad
  \widetilde{\zeta}^\alpha = \frac{1}{\sqrt{s}} \left( \begin{array}{c}
  1 \\ 0
  \end{array} \right)\, ,
\ee
\be\label{eq:Hbgd}
  H = i \kappa\, ,
\ee
where $\kappa$ is a constant. In the frame
\be\label{eq:frame}
  e_1 = d\tau\, ,\quad e_z \equiv e_2 + i e_3 = c\, dz\, ,\quad e_{\bar{z}} \equiv e_2 - i e_3 = c\, d\bar{z},
\ee
with $c$ given by \eqref{eq:def-c}, the other background gauge fields can also be obtained to be
\begin{align}
  V_1 & = \kappa\, ,\label{eq:V1}\\
  V_2 & = 0\, ,\\
  V_3 & = 0\, ,\\
  A_1 & = - \frac{i}{2} \partial_\tau \textrm{log}\, s + \frac{3}{2} \kappa\, ,\label{eq:A1}\\
  A_2 & = \frac{i}{2 c} \partial_z \, \textrm{log} \frac{c}{s} - \frac{i}{2 c} \partial_{\bar{z}} \, \textrm{log} (c s)\, ,\\
  A_3 & = - \frac{1}{2 c} \partial_z \, \textrm{log} \frac{c}{s} - \frac{1}{2 c} \partial_{\bar{z}} \, \textrm{log} (c s)\, ,\label{eq:A3}
\end{align}
where $s (\tau,\, z,\, \bar{z})$ is an arbitrary function which in this paper is chosen to be
\be
  s = e^{- 8 \tau \Phi}\, ,
\ee
where $\Phi$ is a purely imaginary constant,  which will be justified later. With this choice, we can express the background gauge fields \eqref{eq:A1} - \eqref{eq:A3} more explicitly as:
\begin{align}
  A_1 & = 4 i \Phi + \frac{3}{2} \kappa\, ,\\
  A_2 & = \frac{i}{4 L} (z - \bar{z})\, ,\\
  A_3 & = \frac{1}{4 L} (z + \bar{z})\, .
\end{align}

Following the same approach as in \cite{Nian:2013qwa}, we find that the explicit expressions of $A_2$ and $A_3$ will be irrelevant for the following calculations, and the expression showing up in the final result of localization is the following combination:
\be
  A_1 - \frac{1}{2} V_1 + i H = 4 i \Phi\, .
\ee
This a key ingredient in our construction because the nature of the background fields  determines the asymptotic field on the gravity dual.  More precisely, the background fields determine the electric potential and the chemical potential for the angular momentum of the AdS$_4$ black hole.

It  was shown in Ref.~\cite{Nian:2013qwa} that one can introduce an additional rotation of the Killing spinors with a parameter $\Theta$ while keeping the Killing spinor equations \eqref{eq:KillingSpEq-1} - \eqref{eq:KillingSpEq-2} invariant:
  \begin{equation}\label{eq:rotateKilling}
    \zeta \rightarrow e^{i \gamma_1 \Theta} \zeta\, ,\quad \widetilde{\zeta} \rightarrow e^{i \gamma_1 \Theta} \widetilde{\zeta}\, .
  \end{equation}
For the Killing spinor equations \eqref{eq:KillingSpEq-1} - \eqref{eq:KillingSpEq-2} to still hold, we choose
\be
  \Theta = - i \tau \Omega\, ,
\ee
and at the same time we also fix the constant
\be\label{eq:Fixkappa}
  \kappa = - 2 i \Omega\, .
\ee

The two Killing spinor equations \eqref{eq:KillingSpEq-1} and \eqref{eq:KillingSpEq-2} can be combined into one equation:
\be
  (\nabla_\mu - i A_\mu \gamma_1) \xi = - \frac{1}{2} H \gamma_\mu \xi - i V_\mu \gamma_1 \xi + \frac{i}{2} V^\nu \gamma_{\mu\nu} \gamma_1 \xi\, ,
\ee
where a general solution $\xi_\alpha$ to this equation takes the form:
\be
  \xi_\alpha = u \zeta_\alpha + v \widetilde{\zeta}_\alpha =
  \left(\begin{array}{c}
    u\, e^{\tau (\Omega - 4 \Phi)}\\
    v\, e^{- \tau (\Omega - 4 \Phi)}
  \end{array}\right)
\ee
with two complex constants $u$ and $v$. If we require the anti-periodic boundary condition along $S^1$, then the Killing spinor $\xi$ should obey
\be
  \tau \to \tau + L\quad \Rightarrow\quad \xi \to - \xi\, ,
\ee
which leads to the constraint
\be\label{eq:constraint}
  e^{L (\Omega - 4 \Phi)} = -1\quad \Leftrightarrow\quad L (4 \Phi - \Omega) = \pi i\quad (\textrm{mod } 2 \pi i)\, .
\ee
Hence,  the previous assumptions that $\Phi$ and $\Omega$ are purely imaginary constants are justified.  For simplicity, we consider $u = v = 1$. Hence, in the following $\xi$ takes the form:
\be
  \xi_\alpha =
  \left(\begin{array}{c}
    e^{\tau (\Omega - 4 \Phi)}\\
    e^{- \tau (\Omega - 4 \Phi)}
  \end{array}\right)\, .
\ee
Thus, a pair of independent supercharges with anti-periodic boundary conditions can be constructed in the curved space \eqref{eq:metric-1} with complex background fields and we have fulfill the first step for supersymmetric localization.  The amount of supersymmetry in this curved background matches AdS$_4$ black holes in 4d $\mathcal{N}=2$ gauged supergravity \cite{Kostelecky:1995ei}.

As we have seen, the construction of supersymmetry in this subsection is very similar to \cite{Cabo-Bizet:2018ehj} albeit in a different dimension. A similar analysis, relevant for AdS$_7$ black holes, was presented in \cite{Kantor:2019lfo}. The crucial new ingredient in all  cases is the inclusion of  complex background fields and, consequently, spinors with anti-periodic boundary conditions along $S^1$ that do not break supersymmetry completely.

%%%%%%%%%%%%%%%%%%%%%%%%%%%%%%%%%%%%%%%%%%%
%%%%%%%%%%%%%%%%%%%%%%%%%%%%%%%%%%%%%%%%%%%
\subsection{Review of the 3d $\mathcal{N}=2$ SUSY Theories}\label{sec:3dSUSYTheory}
%%%%%%%%%%%%%%%%%%%%%%%%%%%%%%%%%%%%%%%%%%%
%%%%%%%%%%%%%%%%%%%%%%%%%%%%%%%%%%%%%%%%%%%

Let us briefly review the 3d SUSY theories  that can be constructed on a large class of curved backgrounds following the implementation of rigid supersymmetry in \cite{Closset:2012ru}. The background discussed in the previous section is a special case in this class and thus we can borrow many of the results obtained in the literature  (see, for example,  \cite{Closset:2012ru} and \cite{Nian:2013qwa}).

The 3D $\mathcal{N}=2$ vector multiplet in the Wess-Zumino gauge transforms in the following way:
  \begin{align}
  \begin{split}\label{eq:fullSUSYgauge}
    \delta a_\mu & = -i (\zeta \gamma_\mu \widetilde{\lambda} + \widetilde{\zeta} \gamma_\mu \lambda)\, ,\\
    \delta \sigma & = -\zeta \widetilde{\lambda} + \widetilde{\zeta} \lambda\, ,\\
    \delta \lambda & = i \zeta (D + \sigma H) - \frac{i}{2} \varepsilon^{\mu\nu\rho} \gamma_\rho \zeta f_{\mu\nu} - \gamma^\mu\, \zeta (i \partial_\mu \sigma - V_\mu \sigma)\, ,\\
    \delta \widetilde{\lambda} & = -i \widetilde{\zeta} (D+\sigma H) - \frac{i}{2} \varepsilon^{\mu\nu\rho} \gamma_\rho \widetilde{\zeta} f_{\mu\nu} + \gamma^\mu\, \widetilde{\zeta} (i\partial_\mu \sigma + V_\mu \sigma)\, ,\\
    \delta D & = D_\mu (\zeta \gamma^\mu \widetilde{\lambda} - \widetilde{\zeta} \gamma^\mu \lambda) - iV_\mu (\zeta \gamma^\mu \widetilde{\lambda} + \widetilde{\zeta} \gamma^\mu \lambda) - H (\zeta \widetilde{\lambda} - \widetilde{\zeta} \lambda) + \zeta [\widetilde{\lambda},\, \sigma] - \widetilde{\zeta} [\lambda,\, \sigma]\, .
  \end{split}
  \end{align}
  The transformations of the chiral  and the anti-chiral multiplets are given by
  \begin{align}
  \begin{split}\label{eq:fullSUSYmatter}
    \delta \phi & = \sqrt{2} \zeta \psi\, ,\\
    \delta \psi & = \sqrt{2} \zeta F - \sqrt{2} i (z - q \sigma - r H) \widetilde{\zeta} \phi - \sqrt{2} i \gamma^\mu \widetilde{\zeta} D_\mu \phi\, ,\\
    \delta F & = \sqrt{2} i (z - q\sigma - (r-2) H) \widetilde{\zeta} \psi + 2i q \phi \widetilde{\zeta} \widetilde{\lambda} - \sqrt{2} i D_\mu (\widetilde{\zeta} \gamma^\mu \psi)\, ,\\
    \delta \widetilde{\phi} & = -\sqrt{2} \widetilde{\zeta} \widetilde{\psi}\, ,\\
    \delta \widetilde{\psi} & = \sqrt{2} \widetilde{\zeta} \widetilde{F} + \sqrt{2} i (z - q\sigma - rH) \zeta \widetilde{\phi} + \sqrt{2} i \gamma^\mu \zeta D_\mu \widetilde{\phi}\, ,\\
    \delta \widetilde{F} & = \sqrt{2} i (z - q\sigma - (r-2) H) \zeta \widetilde{\psi} + 2 i q \widetilde{\phi} \zeta \lambda - \sqrt{2} i D_\mu (\zeta \gamma^\mu \widetilde{\psi})\, ,
  \end{split}
  \end{align}
  where $z$, $r$ and $q$ denote, respectively,  the central charge, the R-charge and the gauge charge for the chiral multiplet  and
  \begin{equation}
    D_\mu \equiv \nabla_\mu - ir (A_\mu - \frac{1}{2} V_\mu) - iz C_\mu - iq [a_\mu,\, \cdot]\, ,
  \end{equation}
  where $C_\mu$ satisfies
  \begin{equation}\label{eq:C_mu}
    V^\mu = -i \varepsilon^{\mu\nu\rho} \partial_\nu C_\rho\, .
  \end{equation}
  The transformation parameters $\zeta$ and $\widetilde{\zeta}$ satisfy the two Killing spinor equations \eqref{eq:KillingSpEq-1} \eqref{eq:KillingSpEq-2} with opposite R-charges respectively. Suppose that $\zeta$ and $\xi$ are two transformation parameters without tilde, and $\widetilde{\zeta}$ and $\widetilde{\xi}$ are two transformation parameters with tilde. It is checked in Ref.~\cite{Closset:2012ru} that the transformations with only parameters with tilde and only parameters without tilde satisfy the algebra:
  \begin{align}\label{eq:SUSYalg}
    \{\delta_\zeta,\, \delta_\xi\} \varphi & = 0\, , \nonumber\\
    \{\delta_{\widetilde{\zeta}},\, \delta_{\widetilde{\xi}}\} \varphi & = 0\, , \nonumber\\
    \{\delta_\zeta,\, \delta_{\widetilde{\zeta}}\} \varphi & = -2i \left(\mathcal{L}'_K \varphi + \zeta \widetilde{\zeta} (z - rH) \varphi \right)\, ,
  \end{align}
  where $\varphi$ denotes an arbitrary field in the theory, and $K^\mu \equiv \zeta \gamma^\mu \widetilde{\zeta}$ is a Killing vector, while $\mathcal{L}'_K$ is a modified Lie derivative with the local R- and $z$-transformation
  \begin{equation}
    \mathcal{L}'_K \varphi \equiv \mathcal{L}_K \varphi - ir K^\mu (A_\mu - \frac{1}{2} V_\mu) \varphi - iz K^\mu C_\mu \varphi\, .
  \end{equation}

  Under these supersymmetry transformations, the following Lagrangians are invariant:
  \begin{enumerate}
    \item Fayet-Iliopoulos term (for $U(1)$-factors of the gauge group):
          \begin{equation}\label{eq:Lag-1}
            \mathscr{L}_{FI} = \xi (D - a_\mu V^\mu - \sigma H)\, .
          \end{equation}
    \item Gauge-Gauge Chern-Simons  Lagrangian:
          \begin{equation}\label{eq:Lag-2}
            \mathscr{L}_{gg} = \textrm{Tr} \left[\frac{k_{gg}}{4\pi} (i\varepsilon^{\mu\nu\rho} a_\mu \partial_\nu a_\rho - 2 D \sigma + 2i \widetilde{\lambda} \lambda) \right]\, .
          \end{equation}
    \item Gauge-$R$ Chern-Simons Lagrangian (for $U(1)$-factors of the gauge group):
          \begin{equation}\label{eq:Lag-3}
            \mathscr{L}_{gr} = \frac{k_{gr}}{2\pi} \left(i\varepsilon^{\mu\nu\rho} a_\mu \partial_\nu (A_\rho - \frac{1}{2} V_\rho) - DH + \frac{1}{4} \sigma (R - 2 V^\mu V_\mu - 2 H^2)\right)\, .
          \end{equation}
    \item Yang-Mills Lagrangian:
          \begin{align}\label{eq:Lag-4}
            \mathscr{L}_{YM} = & \textrm{Tr} \Bigg[ \frac{1}{4e^2} f^{\mu\nu} f_{\mu\nu} + \frac{1}{2e^2} \partial^\mu \sigma \partial_\mu \sigma - \frac{i}{e^2} \widetilde{\lambda} \gamma^\mu (D_\mu + \frac{i}{2} V_\mu) \lambda - \frac{i}{e^2} \widetilde{\lambda} [\sigma,\, \lambda]\nonumber\\
            {} & + \frac{i}{2e^2} \sigma \varepsilon^{\mu\nu\rho} V_\mu f_{\nu\rho} - \frac{1}{2e^2} V^\mu V_\mu \sigma^2 - \frac{1}{2e^2} (D + \sigma H)^2 + \frac{i}{2e^2} H \widetilde{\lambda} \lambda \Bigg]\, .
          \end{align}
     \item Matter Lagrangian:
           \begin{align}\label{eq:Lag-5}
             \mathscr{L}_{\textrm{mat}} = & \mathscr{D}^\mu \widetilde{\phi} \mathscr{D}_\mu \phi - i \widetilde{\psi} \gamma^\mu \mathscr{D}_\mu \psi - \widetilde{F} F + q (D + \sigma H) \widetilde{\phi} \phi - 2 (r - 1) H (z - q\sigma) \widetilde{\phi} \phi \nonumber\\
             {} & \left((z - q\sigma)^2 - \frac{r}{4} R + \frac{1}{2} (r - \frac{1}{2}) V^\mu V_\mu + r (r - \frac{1}{2}) H^2 \right) \widetilde{\phi} \phi \nonumber\\
             {} & \left(z - q\sigma (r - \frac{1}{2}) H \right) i \widetilde{\psi} \psi + \sqrt{2} i q (\widetilde{\phi} \lambda \psi + \phi \widetilde{\lambda} \widetilde{\psi})\, ,
           \end{align}
           where
           \begin{equation}
             \mathscr{D}_\mu \equiv \nabla_\mu - ir (A_\mu - \frac{1}{2} V_\mu) + i r_0 V_\mu - iz C_\mu - iq [a_\mu,\, \cdot]\, .
           \end{equation}
  \end{enumerate}
  In principle we could also add a superpotential term to the theory:
  \begin{equation}
    \int d^2 \theta\, W + \int d^2 \bar{\theta}\, \overline{W}\, ,
  \end{equation}
which is $\delta$-exact. The superpotential $W$ should be gauge invariant and have R-charge $2$, which imposes contraints on the fields and implicitly affects the final result of the partition function. In this paper, for simplicity,  we do not consider a superpotential term.

%%%%%%%%%%%%%%%%%%%%%%%%%%%%%%%%%%%%%%%%%%%
%%%%%%%%%%%%%%%%%%%%%%%%%%%%%%%%%%%%%%%%%%%
\section{Partition Functions of 3d $\mathcal{N} = 2$ SUSY Theories}\label{sec:Localization}
%%%%%%%%%%%%%%%%%%%%%%%%%%%%%%%%%%%%%%%%%%%
%%%%%%%%%%%%%%%%%%%%%%%%%%%%%%%%%%%%%%%%%%%

We localize the 3d theory reviewed in the previous section following the approach introduced in Ref.~\cite{Nian:2013qwa}, we find that similar steps to those in Ref.~\cite{Nian:2013qwa} apply. The partition function can be expressed as
    \begin{equation}\label{eq:PartFct}
      Z = \frac{1}{|\mathcal{W}|} \int d^r \sigma Z_{\textrm{class}} \, Z_{\textrm{chiral}}^{1-\textrm{loop}}\, Z_{\textrm{vec}}^{1-\textrm{loop}}\, ,
    \end{equation}
where $|\mathcal{W}|$ denotes the order of the Weyl group associated to the gauge group.    The classical contribution, $Z_{\textrm{class}}$, the 1-loop determinants $Z_{\textrm{chiral}}^{1-\textrm{loop}}$ for the 3d chiral multiplet and $Z_{\textrm{vec}}^{1-\textrm{loop}}$ for the 3d vector multiplet can be obtained as follows.

%%%%%%%%%%%%%%%%%%%%%%%%%%%%%%%%%%%%%%%%%%%
\subsection{Saddle-Point Configurations}\label{sec:SaddlePt}

Following the standard approach of supersymmetric localization, we can deform the original theory by adding to the Lagrangian a $\delta$-exact  term $t\, \delta \mathcal{V}$, where $\mathcal{V}$ is chosen to be
\be
  \mathcal{V} = \psi^\dagger \delta \psi + \widetilde{\psi}^\dagger \delta \widetilde{\psi} + \lambda^\dagger \delta \lambda + \widetilde{\lambda}^\dagger \delta \widetilde{\lambda}\, .
\ee
The localization locus follows from 
    \begin{equation}\label{eq:LocalCond}
      \delta \psi = 0\, ,\quad \delta \widetilde{\psi} = 0\, ,\quad \delta \lambda = 0\, ,\quad \delta \widetilde{\lambda} = 0\, ,
    \end{equation}
which can be directly  obtained from the supersymmetry transformations \eqref{eq:fullSUSYgauge} and \eqref{eq:fullSUSYmatter}. These BPS equations can be written more precisely as follows:
\begin{align}
  \sqrt{2} \zeta F - \sqrt{2} i (z - q \sigma - r H) \widetilde{\zeta} \phi - \sqrt{2} i \gamma^\mu \widetilde{\zeta} D_\mu \phi & = 0\, ,\\
  \sqrt{2} \widetilde{\zeta} \widetilde{F} + \sqrt{2} i (z - q\sigma - rH) \zeta \widetilde{\phi} + \sqrt{2} i \gamma^\mu \zeta D_\mu \widetilde{\phi} & = 0\, ,\\
  i \zeta (D + \sigma H) - \frac{i}{2} \varepsilon^{\mu\nu\rho} \gamma_\rho \zeta f_{\mu\nu} - \gamma^\mu\, \zeta (i \partial_\mu \sigma - V_\mu \sigma) & = 0\, ,\\
  -i \widetilde{\zeta} (D+\sigma H) - \frac{i}{2} \varepsilon^{\mu\nu\rho} \gamma_\rho \widetilde{\zeta} f_{\mu\nu} + \gamma^\mu\, \widetilde{\zeta} (i\partial_\mu \sigma + V_\mu \sigma) & = 0\, .
\end{align}
We can multiply these equations with $\zeta$ and $\widetilde{\zeta}$ from the left and simplify them using the spinor bilinears. The resulting equations are some scalar-valued partial differential equations, and the solutions are
\be
  \phi = F = \widetilde{\phi} = \widetilde{F} = 0\, ,
\ee
\be\label{eq:sigmaClass}
  \sigma = \frac{i f_{23}}{V_1} = - \frac{\mathfrak{m}}{4 \Omega L^2}\, ,\quad D = - \sigma H = \frac{\mathfrak{m}}{2 L^2}\, ,
\ee
\be
  a_1 = a\, ,\quad f_{23} = \frac{\mathfrak{m}}{2 L^2}\, ,
\ee
while all the other fields vanishing. Among these solutions, the dynamical gauge fields $a_\mu$ have the saddle-point configurations:
\be
  a_1 = a\, ,\quad a_2 = \frac{i \mathfrak{m} (z - \bar{z})\, e^{- \Omega \tau}}{4 L}\, ,\quad a_3 = \frac{\mathfrak{m} (z + \bar{z})\, e^{- \Omega \tau}}{4 L}\, ,
\ee
or equivalently, in the frame \eqref{eq:frame} given by
\be
  a_\tau = a\, ,\quad a_z = - \frac{i \mathfrak{m} \bar{z}\, e^{- \Omega \tau}}{2 \left(e^{\Omega \tau} + |z|^2 e^{- \Omega \tau}\right)}\, ,\quad a_{\bar{z}} = \frac{i \mathfrak{m} z\, e^{- \Omega \tau}}{2 \left(e^{\Omega \tau} + |z|^2 e^{- \Omega \tau} \right)}\, ,
\ee
where $a$ and $\mathfrak{m}$ are constants, and $\mathfrak{m}$ takes value in the Cartan subalgebra of the gauge group, such that
\be
  \rho (\mathfrak{m}) \in \mathbb{Z}\, ,\quad \alpha (\mathfrak{m}) \in \mathbb{Z}
\ee
with the weight vector $\rho$ in the representation $R$ and the root vector $\alpha$ in the adjoint representation. The field strength $f_{23}$ leads to the quantization condition:
\be
  \frac{1}{2 \pi} \int f_{23}\, e^2 \wedge e^3 = \mathfrak{m}\, ,
\ee
which clarifies explicitly why this is a monopole-type configuration.

%%%%%%%%%%%%%%%%%%%%%%%%%%%%%%%%%%%%%%
\subsection{Classical Contribution}

The classical part of the partition function, $Z_{\textrm{class}}$, includes the contributions from $\mathscr{L}_{FI}$, $\mathscr{L}_{gg}$ and $\mathscr{L}_{gr}$. In this paper, we only turn on $\mathscr{L}_{gg}$, which based on \eqref{eq:Hbgd} \eqref{eq:Fixkappa} \eqref{eq:sigmaClass} can be more explicitly expressed as
\be\label{eq:Z_CS}
  Z_{CS,\, \mathfrak{m}} = \textrm{exp} \left(i \int d^3 x \sqrt{g} \mathscr{L}_{gg} \right) = \textrm{exp} \left(\frac{k_{gg}}{4 \pi} \int d^3 x \sqrt{g}\, \sum_i \Big[\frac{\mathfrak{m}_i^2}{4 \Omega L^4} + \frac{i a_i \mathfrak{m}_i}{2 L^2} \Big] \right)\, .
\ee
Note that we have the standard Chern-Simons contribution of the type $a\wedge F$ which is linear in the magnetic flux $\mathfrak{m}_i$ but we also have a contribution, coming from the term $D\sigma$ which is quadratic in the flux.

There are two subtle points that we would like to emphasize. First, for the special curved space given by the metric \eqref{eq:metric-1} we have to carefully consider the quantization condition of the Chern-Simons level $k$. As discussed in e.g. \cite{Tong:2016kpv}, for a periodic Euclidean time $S^1$ the electron wave function transforms as $e^{i e \omega / \hbar}$, which induces a gauge transformation:
\be
  A_\mu \rightarrow A_\mu + \partial_\mu \omega\, .
\ee
Consequently, the Chern-Simons level has to satisfy the condition:
\be
  \frac{\hbar k}{e^2} \in \mathbb{Z}\, .
\ee
In fact, the additional shifts $\sim \kappa$ in the background gauge fields $V_1$ \eqref{eq:V1} and $A_1$ \eqref{eq:A1} can be understood in this way, and the additional rotation of the Killing spinor \eqref{eq:rotateKilling} implies that the wave function of physical spinors along a periodic $S^1$ should be $e^{- \tau \Omega L / \pi \ell}$, with $\ell$ denoting the effective circumference of $S^1$. In this case, the corresponding Chern-Simons level obeys
\be
  \frac{i k_{gg} \Omega L}{2 \pi^2} \in \mathbb{Z}\, .
\ee
For later convenience, let us define an integer-valued new constant $k$ as
\be
  k \equiv \frac{i k_{gg} \Omega L}{2 \pi^2}\, .
\ee
Therefore,
\be
  \frac{k_{gg}}{4 \pi} = \frac{- i \pi k}{2 \Omega L}\, .
\ee

The second subtle point is that similar to the case of rotating BTZ black holes as quotients of $\mathbb{H}_3$ \cite{Keeler:2018lza}, we should require a modified periodicity condition in order that the coordinates are regular for $\varphi \in [0,\, 2 \pi]$:
\be
  \varphi \sim \varphi + 2 \pi\, ,\quad \tau \sim \tau - \frac{2 \pi i}{\Omega}\, .
\ee
Consequently, it will modify the Chern-Simons term by changing the effective circumference of $S^1$, and the classical contribution from \eqref{eq:Z_CS} becomes
\be\label{eq:Z_CS-2}
  Z_{CS,\, \mathfrak{m}} = \textrm{exp} \left(\frac{4 \pi^3 k L}{\Omega^2} \sum_i \Big[\frac{\mathfrak{m}_i^2}{4 \Omega L^4} + \frac{i a_i \mathfrak{m}_i}{2 L^2} \Big] \right)\, .
\ee

\subsection{1-Loop Determinants}

To compute the 1-loop determinants in supersymmetric localization there are various approaches, for instance, using index theorem or explicit mode expansions. In this paper, we follow the approach discussed in \cite{Hama:2011ea, Alday:2013lba, Nian:2013qwa} by considering the unpaired spinors and the missing spinors, whose contributions to the 1-loop determinants are not canceled by their superpartners. Although \cite{Nian:2013qwa} focuses on the squashed $S^3$ as the curved space, the general results there work for all the curved spaces in the class defined by the metric \eqref{eq:metric-2} and the frame \eqref{eq:frame}. Hence, in this subsection, we borrow the general results from \cite{Nian:2013qwa} and apply them to the case \eqref{eq:metric-1} studied in this paper. For more details, we refer to \cite{Nian:2013qwa}.

For the 3d chiral multiplet, the 1-loop determinant $Z_{\textrm{chiral}}^{1-\textrm{loop}}$ is expressed as the quotient of the eigenvalues of the fermionic and the bosonic operators, $\hat{M}_\psi$ and $\hat{M}_\phi$, given by
    \begin{align}
      \hat{M}_\psi & = i e^{-2 \textrm{Im} \Theta} \left[\nabla_1 - i(r-2) (A_1 - \frac{1}{2} V_1) - i(z - q\sigma) C_1 - (z - q\sigma) + (r-2) H \right] \, ,\\
      \hat{M}_\phi & = i e^{-2 \textrm{Im} \Theta} \left[\nabla_1 - ir (A_1 - \frac{1}{2} V_1) - i(z - q\sigma) C_1 + (\bar{z} - q\bar{\sigma}) + rH \right] \, .
    \end{align}
where we can use the background gauge symmetry of \eqref{eq:C_mu} to set $C_1 = 0$, and we assume $z$ and $\sigma$ to be purely imaginary. For simplicity, we also set the central charge to $z = 0$ in the following calculations.

In the frame \eqref{eq:frame}, the covariant derivative $\nabla_1$ can be expressed as
\be\label{eq:nabla_1}
  \nabla_1 = \nabla_{\widetilde{\tau}} - i \Omega \nabla_{\widetilde{\varphi}}\, .
\ee
Taking into account the eigenmodes of $\nabla_{\widetilde{\tau}}$ and $\nabla_{\widetilde{\varphi}}$ given by $2 \pi i n_0 / L$ and $2 i m$, with the winding modes on $S^1$ labelled by $n_0 \in \mathbb{Z}$ and the spherical harmonics on $S^2$ labelled by $(j, m)$ respectively, one can straightforwardly obtain the 1-loop determinant $Z_{\textrm{chiral}}^{1-\textrm{loop}}$:
\begin{align}
  Z_{\textrm{chiral}}^{1-\textrm{loop}} & = \prod_{\rho \in R} \prod_{n_0 \in \mathbb{Z}} \prod_{j \in \frac{1}{2} \mathbb{N}} \prod_{m = -j}^j \frac{2 \pi i n_0 - i L \rho (a_1) + 2 L \Omega m + L (r - 2) 4 \Phi + L \rho (\sigma)}{2 \pi i n_0 - i L \rho (a_1) + 2 L \Omega m + L r 4 \Phi + L \rho (\sigma)} \nonumber\\
  {} & = \prod_{\rho \in R} \prod_{n_0 \in \mathbb{Z}} \prod_{j \in \frac{1}{2} \mathbb{N}} \prod_{m = -j}^j \frac{2 \pi i n_0 - i L \rho (a) + 2 L \Omega m + L r 4 \Phi - 2 \Omega L - \frac{\rho (\mathfrak{m})}{4 \Omega L}}{2 \pi i n_0 - i L \rho (a) + 2 L \Omega m + L r 4 \Phi - \frac{\rho (\mathfrak{m})}{4 \Omega L}} \nonumber\\
  {} & = \prod_{\rho \in R} \prod_{n_0 \in \mathbb{Z}} \prod_{j \in \frac{1}{2} \mathbb{N}} \prod_{m = -j}^j \frac{2 \pi i n_0 - i L \rho (a) + 2 L \Omega (m - 1) + L r 4 \Phi - \frac{\rho (\mathfrak{m})}{4 \Omega L}}{2 \pi i n_0 - i L \rho (a) + 2 L \Omega m + L r 4 \Phi - \frac{\rho (\mathfrak{m})}{4 \Omega L}} \nonumber\\
  {} & = \prod_{\rho \in R} \prod_{n_0 \in \mathbb{Z}} \prod_{j \in \frac{1}{2} \mathbb{N}} \frac{2 \pi i n_0 - i L \rho (a) + 2 L \Omega (- j - 1) + L r 4 \Phi - \frac{\rho (\mathfrak{m})}{4 \Omega L}}{2 \pi i n_0 - i L \rho (a) + 2 L \Omega j + L r 4 \Phi - \frac{\rho (\mathfrak{m})}{4 \Omega L}} \nonumber\\
  {} & = \prod_{\rho \in R} \prod_{n_0 \in \mathbb{Z}} \prod_{n_1 \in \mathbb{N}} \frac{2 \pi i n_0 - i L \rho (a) - L \Omega n_1 - 2 L \Omega + L r 4 \Phi - \frac{\rho (\mathfrak{m})}{4 \Omega L}}{2 \pi i n_0 - i L \rho (a) + L \Omega n_1 + L r 4 \Phi - \frac{\rho (\mathfrak{m})}{4 \Omega L}} \nonumber\\
  {} & = \prod_{\rho \in R} \prod_{n_0 \in \mathbb{Z}} \prod_{n_1 \in \mathbb{N}} \frac{2 \pi i n_0 - i L \rho (a) - L \Omega n_1 + L (r - 2) 4 \Phi - \frac{\rho (\mathfrak{m})}{4 \Omega L}}{2 \pi i n_0 - i L \rho (a) + L \Omega n_1 + L r 4 \Phi - \frac{\rho (\mathfrak{m})}{4 \Omega L}} \, ,\label{eq:1-loopChiralProduct}
\end{align}
where $n_1 \equiv 2 j$, while $\rho$ denotes the weight vector in a representation $R$ of the gauge group, and we have used the condition \eqref{eq:constraint}.

For the 3d vector multiplet, we first choose the gauge $\nabla^\mu A_\mu = 0$. Correspondingly, we add a gauge fixing term to the Lagrangian:
    \begin{equation}
      \mathscr{L}_{gf} = \textrm{Tr} \left[\bar{c} \nabla^\mu \nabla_\mu c + b \nabla^\mu a_\mu \right]\, .
    \end{equation}
After incorporating  the cancellation among various modes, it was shown in \cite{Nian:2013qwa} that the 1-loop determinant $Z_{\textrm{vec}}^{1-\textrm{loop}}$ can be expressed as the quotient of the eigenvalues of the fermionic and the bosonic operators,  $M_\Phi$ and $M_B$ given by
\begin{align}
  M_\Phi & = i \Big[\alpha(\sigma) + \nabla_1\Big]\, ,\\
  M_B & = i \Big[\alpha(\sigma) + \nabla_1\Big]\, ,
\end{align}
where $\Phi$ and $B$ are spin-0 and spin-1 modes respectively. Similar to the chiral multiplet, we can also use \eqref{eq:nabla_1} to express the eigenmodes of $\nabla_1$ into the eigenmodes of $\nabla_{\widetilde{\tau}}$ and $\nabla_{\widetilde{\varphi}}$ given by $2 \pi i n_0 / L$ and $2 i m$, with the winding modes on $S^1$ labelled by $n_0 \in \mathbb{Z}$ and the spherical harmonics on $S^2$ labelled by $(j, m)$ respectively. Consequently, we obtain 1-loop determinant $Z_{\textrm{vec}}^{1-\textrm{loop}}$:

\begin{align}
  Z_{\textrm{vec}}^{1-\textrm{loop}} & = \prod_{\alpha \in \textrm{Adj}} \prod_{n_0 \in \mathbb{Z}} \prod_{j \in \frac{1}{2} \mathbb{N}} \prod_{m = -j}^j \frac{2 \pi i n_0 - i L \alpha (a_1) + 2 L \Omega m + L \alpha (\sigma)}{2 \pi i n_0 - i L \alpha (a_1) + 2 L \Omega (m - 1) + L \alpha (\sigma)} \nonumber\\
  {} & = \prod_{\alpha \in \textrm{Adj}} \prod_{n_0 \in \mathbb{Z}} \prod_{j \in \frac{1}{2} \mathbb{N}} \frac{2 \pi i n_0 - i L \alpha (a) + 2 L \Omega j - \frac{\alpha (\mathfrak{m})}{4 \Omega L}}{2 \pi i n_0 - i L \alpha (a) + 2 L \Omega (- j - 1) - \frac{\alpha (\mathfrak{m})}{4 \Omega L}} \nonumber\\
  {} & = \prod_{\alpha \in \textrm{Adj}} \prod_{n_0 \in \mathbb{Z}} \prod_{n_1 \in \mathbb{N}} \frac{2 \pi i n_0 - i L \alpha (a) + L \Omega n_1 - \frac{\alpha (\mathfrak{m})}{4 \Omega L}}{2 \pi i n_0 - i L \alpha (a) - L \Omega n_1 - 2 L \Omega - \frac{\alpha (\mathfrak{m})}{4 \Omega L}} \nonumber\\
  {} & = \prod_{\alpha \in \textrm{Adj}} \prod_{n_0 \in \mathbb{Z}} \prod_{n_1 \in \mathbb{N}} \frac{2 \pi i n_0 - i L \alpha (a) + L \Omega n_1 - \frac{\alpha (\mathfrak{m})}{4 \Omega L}}{2 \pi i n_0 - i L \alpha (a) - L \Omega n_1 - 8 L \Phi - \frac{\alpha (\mathfrak{m})}{4 \Omega L}} \nonumber\\
  {} & = \prod_{\alpha \in \textrm{Adj}} \prod_{n_0 \in \mathbb{Z}} \prod_{n_1 \in \mathbb{N}} \frac{- 2 \pi i n_0 + i L \alpha (a) + L \Omega n_1 + \frac{\alpha (\mathfrak{m})}{4 \Omega L}}{- 2 \pi i n_0 + i L \alpha (a) - L \Omega n_1 - 8 L \Phi + \frac{\alpha (\mathfrak{m})}{4 \Omega L}} \nonumber\\
  {} & = \prod_{\alpha \in \textrm{Adj}} \prod_{n_0 \in \mathbb{Z}} \prod_{n_1 \in \mathbb{N}} \frac{2 \pi i n_0 - i L \alpha (a) - L \Omega n_1 - \frac{\alpha (\mathfrak{m})}{4 \Omega L}}{2 \pi i n_0 - i L \alpha (a) + L \Omega n_1 + 8 L \Phi - \frac{\alpha (\mathfrak{m})}{4 \Omega L}} \, ,\label{eq:1-loopVecProduct}
\end{align}
where $n_1 \equiv 2 j$, while $\alpha$ denotes the root vector in the adjoint representation of the gauge group, and we have used the condition \eqref{eq:constraint} as well as the following reflection symmetry of the expression:
\be
  n_0 \rightarrow - n_0\, ,\quad \alpha \rightarrow - \alpha\, .
\ee
Hence, we see that the 1-loop determinant of the vector multiplet can be viewed as the one of an adjoint chiral multiplet with the R-charge $r = 2$. Of course, in the above expression we have omitted a term that eventually cancels  with the integration measure following \cite{Martelli:2011fu, Alday:2013lba, Assel:2014paa, Cabo-Bizet:2018ehj}  which elaborated on the original treatment in \cite{Kapustin:2009kz}.

Similar to Ref.~\cite{Cabo-Bizet:2018ehj}, we can regularize the infinite products of the 1-loop determinants using double gamma functions. After a few steps, we find that the infinite products \eqref{eq:1-loopChiralProduct} and \eqref{eq:1-loopVecProduct} can be regularized as follows:
\begin{align}
  Z_{\textrm{chiral}}^{1-\textrm{loop}} & = \prod_{\rho \in R} \frac{\Gamma_2 (\gamma + \delta | 1, \beta)\cdot \Gamma_2 (1 - \gamma - \delta | 1, - \beta)}{\Gamma_2 (\gamma - \delta | 1, \beta)\cdot \Gamma_2 (1 - \gamma + \delta | 1, - \beta)}\, ,\label{eq:1-loopChiralRegTemp}\\
  Z_{\textrm{vec}}^{1-\textrm{loop}} & = \left[Z_{\textrm{chiral}}^{1-\textrm{loop}}\right]_{\rho = \alpha \in \textrm{Adj},\, r=2}\, ,\label{eq:1-loopVecRegTemp}
\end{align}
where $\Gamma_2 (x | a_1, a_2)$ is the double gamma function, and
\be\label{eq:DefParam-1}
  \beta \equiv \frac{i L \Omega}{2 \pi}\, ,\quad \gamma \equiv \frac{2 i L \Phi}{\pi}\, ,
\ee
\be\label{eq:DefParam-2}
  \delta \equiv \frac{1}{2 \pi} \left[- \frac{i\, \rho (\mathfrak{m})}{4 \Omega L} + L \rho (a) - i (r - 1) 4 L \Phi \right]\, .
\ee
Due to the constraint \eqref{eq:constraint}, the parameters $\beta$ and $\gamma$ are not independent, instead they satisfy
\be
  \gamma = \beta + \frac{2 n + 1}{2}\quad (n \in \mathbb{Z})\, .
\ee
Using the identity \eqref{eq:Gamma2Id}, we can further rewrite the expression \eqref{eq:1-loopChiralRegTemp} as
\begin{align}
  Z_{\textrm{chiral}}^{1-\textrm{loop}} & = \prod_{\rho \in R} e^{- \pi i \big[\zeta_2 (0, \gamma + \delta | 1, \beta) - \zeta_2 (0, \gamma - \delta | 1, \beta) \big]}\, \prod_{k \in \mathbb{N}} \frac{1 - e^{2 \pi i (\gamma - \delta + k \beta)}}{1 - e^{2 \pi i (\gamma + \delta + k \beta)}}\nonumber\\
  {} & \equiv \prod_{\rho \in R} e^{- \pi i \Psi} \frac{\big(e^{2 \pi i (\gamma - \delta)};\, e^{2 \pi i \beta} \big)_\infty}{\big(e^{2 \pi i (\gamma + \delta)};\, e^{2 \pi i \beta} \big)_\infty}\, ,\label{eq:1-loopChiralReg}
\end{align}
where
\be
  \Psi \equiv - \frac{\delta}{\beta} (1 + \beta - 2 \gamma) = \delta \left(1 + \frac{2 n}{\beta} \right)\quad (n \in \mathbb{Z})\, ,
\ee
and $(a;\, q)_m$ denotes the $q$-Pochhammer symbol defined as
\be
  (a;\, q)_m \equiv \prod_{k=0}^{m-1} (1 - a\, q^k)\, .
\ee
Similarly, the 1-loop determinant for the vector multiplet can be further regularized as
\begin{align}
  Z_{\textrm{vec}}^{1-\textrm{loop}} & = \left[Z_{\textrm{chiral}}^{1-\textrm{loop}}\right]_{\rho = \alpha \in \textrm{Adj},\, r=2} \nonumber\\
  {} & = \prod_{\alpha \in \textrm{Adj}} \Bigg[e^{- \pi i \Psi} \frac{\big(e^{2 \pi i (\gamma - \delta)};\, e^{2 \pi i \beta} \big)_\infty}{\big(e^{2 \pi i (\gamma + \delta)};\, e^{2 \pi i \beta} \big)_\infty}\Bigg]_{\rho = \alpha,\, r = 2}\, .\label{eq:1-loopVecReg}
\end{align}

\subsection{Full Partition Function}

To summarize, we can consider an arbitrary 3d $\mathcal{N}=2$ supersymmetric gauge theory defined on the curved background \eqref{eq:metric-1}. Using the supersymmetric localization, we obtain the exact partition function of the theory given by \eqref{eq:PartFct}:
\begin{displaymath}
  Z = \frac{1}{|\mathcal{W}|} \sum_{\mathfrak{m}} \int d^r a Z_{\textrm{class}} \, Z_{\textrm{chiral}}^{1-\textrm{loop}}\, Z_{\textrm{vec}}^{1-\textrm{loop}}\, ,
\end{displaymath}
where the classical contribution $Z_{\textrm{class}}$ is given only by the Chern-Simons term $Z_{CS,\, \mathfrak{m}}$ \eqref{eq:Z_CS-2}, while the regularized 1-loop determinants $Z_{\textrm{chiral}}^{1-\textrm{loop}}$ and $Z_{\textrm{vec}}^{1-\textrm{loop}}$ are given by \eqref{eq:1-loopChiralReg} and \eqref{eq:1-loopVecReg}.

\subsection{ABJM Theory}

We now consider the ABJM theory as a special example of the general case discussed in this section. The ABJM theory is a 3d gauge theory with gauge group $U(N) \times U(N)$, whose Chern-Simons levels are $k$ and $-k$, respectively \cite{Aharony:2008ug}. This theory describes the low-energy effective theory of $N$ M2-branes probing $\mathbb{C}^4 / \mathbb{Z}_k$. The R-symmetry group is $SO(6) \cong SU(4)$. In 3d $\mathcal{N}=2$ language, the field content includes 2 vector multiplets of R-charge $2$ with the gauge fields $A_\mu$ and $\widetilde{A}_\mu$ for $U(N) \times U(N)$ respectively, while the matter fields are
\be
  C_I = (A_1,\, A_2,\, \bar{B}^{\dot{1}},\, \bar{B}^{\dot{2}})\, ,\quad \Psi^I = (-\psi_2,\, \psi_1,\, - \bar{\chi}^{\dot{2}},\, \bar{\chi}^{\dot{1}})\, ,
\ee
which can be group into 4 chiral multiplets of R-charge $\frac{1}{2}$ as
\be
  (A_a, \psi_{a \alpha}) \in ({\bf{N}, \bar{\bf{N}}})\, ,\quad (B_{\dot{a}}, \chi_{\dot{a} \alpha}) \in (\bar{\bf{N}}, \bf{N})\, ,
\ee
where $a = 1, 2$ and $\dot{a} = \dot{1}, \dot{2}$ are double indices of $SU(2) \times SU(2) \subset SU(4)_R$. We use $h_1$, $h_2$ and $h_3$ to denote the three Cartans of $SO(6)_R$, while $\frac{1}{2} (h_1 \pm h_2)$ denote the two Cartans of the subgroup $SU(2) \times SU(2)$. The charges $h_{1, 2, 3}$ of different fields are listed in the following table \cite{Bhattacharya:2008bja, Kim:2009wb}:
\begin{table}[h!]
\centering
\begin{tabular}{c|ccc}
\hline
fields & $h_1$ & $h_2$ & $h_3$\\
\hline
$(A_1,\, A_2)$ & $\left(\frac{1}{2},\, - \frac{1}{2}\right)$ & $\left(\frac{1}{2},\, - \frac{1}{2}\right)$ & $\left(- \frac{1}{2},\, - \frac{1}{2}\right)$ \\
$(B_{\dot{1}},\, B_{\dot{2}})$ & $\left(\frac{1}{2},\, - \frac{1}{2}\right)$ & $\left(- \frac{1}{2},\, \frac{1}{2}\right)$ & $\left(- \frac{1}{2},\, - \frac{1}{2}\right)$ \\
$(\psi_{1 \pm},\, \psi_{2 \pm})$ & $\left(\frac{1}{2},\, - \frac{1}{2}\right)$ & $\left(\frac{1}{2},\, - \frac{1}{2}\right)$ & $\left(\frac{1}{2},\, \frac{1}{2}\right)$ \\
$(\chi_{\dot{1} \pm},\, \chi_{\dot{2} \pm})$ & $\left(\frac{1}{2},\, - \frac{1}{2}\right)$ & $\left(- \frac{1}{2},\, \frac{1}{2}\right)$ & $\left(\frac{1}{2},\, \frac{1}{2}\right)$ \\
\hline
$A_\mu,\, \widetilde{A}_\mu$ & $0$ & $0$ & $0$ \\
$\lambda_\pm,\, \widetilde{\lambda}_\pm$ & $0$ & $0$ & $-1$ \\
$\sigma,\, \widetilde{\sigma}$ & $0$ & $0$ & $0$ \\
\hline
$Q_\pm$ & $0$ & $0$ & $1$ \\
$S^\pm$ & $0$ & $0$ & $-1$
\end{tabular}
\caption{Charges $(h_1,\, h_2,\, h_3)$ of the fields}
\end{table}

The partition function for the ABJM theory on the background \eqref{eq:metric-1} is given by
\begin{align}
  Z & = \frac{1}{\left(N ! \right)^2} \sum_{\mathfrak{m}_i,\, \widetilde{\mathfrak{m}}_i} \int \prod_{i=1}^N \frac{d\lambda_i\, d\widetilde{\lambda}_i}{(2 \pi)^2}\nonumber\\
  {} & \qquad\qquad \cdot e^{\frac{k_{gg}}{4 \pi} \int d^3 x \sqrt{g}\, \sum_i \left(\frac{\mathfrak{m}_i^2}{4 \Omega L^4} + \frac{i a_i \mathfrak{m}_i}{2 L^2} - \frac{\widetilde{\mathfrak{m}}_i^2}{4 \Omega L^4} - \frac{i \widetilde{a}_i \widetilde{\mathfrak{m}}_i}{2 L^2}\right)} \cdot e^{- \pi i \sum_{i, j} \sum_{I=1}^4 \Psi_I^c} \cdot e^{- \pi i \sum_{i \neq j} \sum_{K=1}^2 \Psi_K^v} \nonumber\\
  {} & \qquad\qquad \cdot \left(\prod_{i, j} \prod_{I=1}^4 \frac{\big(e^{2 \pi i (\gamma - \delta_I^c)};\, e^{2 \pi i \beta} \big)_\infty}{\big(e^{2 \pi i (\gamma + \delta_I^c)};\, e^{2 \pi i \beta} \big)_\infty} \right) \cdot \left(\prod_{i \neq j} \prod_{K=1}^2 \frac{\big(e^{2 \pi i (\gamma - \delta_K^v)};\, e^{2 \pi i \beta} \big)_\infty}{\big(e^{2 \pi i (\gamma + \delta_K^v)};\, e^{2 \pi i \beta} \big)_\infty} \right)\, ,\label{eq:ABJMmatrixModel}
\end{align}
where we define new variables $\lambda_i$ and $\widetilde{\lambda}_i$:
\be
  i \frac{\lambda_i}{2 \pi} \equiv - \frac{i \mathfrak{m}_i}{4 \Omega L} + L a_i\, ,\quad i \frac{\widetilde{\lambda}_i}{2 \pi} \equiv - \frac{i \widetilde{\mathfrak{m}}_i}{4 \Omega L} + L \widetilde{a}_i\, .
\ee
 Here $a_i$ and $\widetilde{a}_i$ denote the values of $a$ and $\widetilde{a}$ in the $i$-th Cartan of the gauge group, which should not be confused with the saddle-point configurations of $a_\mu$ discussed in \ref{sec:SaddlePt}. The explicit expressions of $\delta$'s for the 4 chiral multiplets ($I = 1,\, \cdots,\, 4$) in the bifundamental representation and the 2 vector multiplets ($K = 1,\, 2$) in the adjoint representation are
\begin{align}
  \delta_{I = 1, 2}^c & \equiv \frac{1}{2 \pi} \Bigg[i \left(\frac{\lambda_i - \widetilde{\lambda}_j}{2 \pi} \right) + 2 i L \Phi \pm \frac{1}{2} \varphi_1 \Bigg]\, ,\label{eq:DefDelta-1}\\
  \delta_{I = 3, 4}^c & \equiv \frac{1}{2 \pi} \Bigg[i \left(\frac{\widetilde{\lambda}_j - \lambda_i}{2 \pi} \right) + 2 i L \Phi \pm \frac{1}{2} \varphi_2 \Bigg]\, ,\\
  \delta_{K = 1}^v & \equiv \frac{1}{2 \pi} \Bigg[i \left(\frac{\lambda_i - \lambda_j}{2 \pi} \right) - 4 i L \Phi \Bigg]\, ,\\
  \delta_{K = 2}^v & \equiv \frac{1}{2 \pi} \Bigg[i \left(\frac{\widetilde{\lambda}_i - \widetilde{\lambda}_j}{2 \pi} \right) - 4 i L \Phi \Bigg]\, ,\label{eq:DefDelta-4}
\end{align}
where $\varphi_1$ and $\varphi_2$ denote the chemical potentials corresponding to the Cartans $\frac{1}{2} (h_1 \pm h_2)$ of the subgroup $SU(2) \times SU(2)$ of the R-symmetry group. In the second line of \eqref{eq:ABJMmatrixModel}, $\Psi_I^c$ ($I = 1,\, \cdots,\, 4$) and $\Psi_K^v$ ($K = 1,\, 2$) are related to $\delta_I^c$ and $\delta_K^v$ as
\be
  \Psi_I^c = \delta_I^c \left(- \frac{1}{\beta} \right) (1 + \beta - 2 \gamma)\, ,\quad \Psi_K^v = \delta_K^v \left(- \frac{1}{\beta} \right) (1 + \beta - 2 \gamma)\, .
\ee
Due to the identity
\be
  \sum_{i, j} \sum_{I = 1}^4 \Psi_I^c - \sum_{i \neq j} \sum_{K = 1}^2 \Psi_K^v = 0\, ,
\ee
the second line of Eq.~\eqref{eq:ABJMmatrixModel} will become $1$, which is consistent with the fact that there is no conformal anomaly in 3d.

We can work with the new variables $\lambda_i$ and $\widetilde{\lambda}_i$, then the magnetic fluxes $\mathfrak{m}_i$ and $\widetilde{\mathfrak{m}}_i$ appear only in the classical part of the partition function, more precisely, the gauge-gauge Chern-Simons term. Let us take a closer look at this term. In terms of $\lambda_i$ and $\widetilde{\lambda}_i$, this term can be expressed as follows:
\begin{align}
  Z_{CS} & = \sum_{\mathfrak{m}_i,\, \widetilde{\mathfrak{m}}_i} e^{\frac{4 \pi^3 k L}{\Omega^2} \sum_i \left(\frac{\mathfrak{m}_i^2}{4 \Omega L^4} + \frac{i a_i \mathfrak{m}_i}{2 L^2} - \frac{\widetilde{\mathfrak{m}}_i^2}{4 \Omega L^4} - \frac{i \widetilde{a}_i \widetilde{\mathfrak{m}}_i}{2 L^2}\right)} \nonumber\\
  {} & = \sum_{\mathfrak{m}_i,\, \widetilde{\mathfrak{m}}_i} e^{\sum_i \left(\frac{k \pi^3 \mathfrak{m}_i^2}{2 L^3 \Omega^3} - \frac{k \pi^2 \mathfrak{m}_i \lambda_i}{L^2 \Omega^2} - \frac{k \pi^3 \widetilde{\mathfrak{m}}_i^2}{2 L^3 \Omega^3} + \frac{k \pi^2 \widetilde{\mathfrak{m}}_i \widetilde{\lambda}_i}{L^2 \Omega^2} \right)}\, .\label{eq:Z_CS-new}
\end{align}
Applying the Poisson resummation formula:
\be
  \sum_{m = - \infty}^\infty e^{- \pi m^2 A + 2 \pi m A s} = \frac{e^{\pi A s^2}}{\sqrt{A}} \sum_{n = - \infty}^\infty e^{- \frac{\pi n^2}{A} - 2 \pi i n s}\, ,
\ee
we obtain from \eqref{eq:Z_CS-new} the following expression:
\be\label{eq:Z_CS-final}
  Z_{CS} = e^{\sum_i \left(- \frac{k \pi \lambda_i^2}{2 L \Omega} + \frac{k \pi \widetilde{\lambda}_i^2}{2 L \Omega}\right)} \sum_{\mathfrak{n}_i,\, \widetilde{\mathfrak{n}}_i} e^{\sum_i \left[\frac{2 L^3 \Omega^3 \mathfrak{n}_i^2}{\pi k} - 2 i L \Omega \mathfrak{n}_i \lambda_i - \frac{2 L^3 \Omega^3 \widetilde{\mathfrak{n}}_i^2}{\pi k} + 2 i L \Omega \widetilde{\mathfrak{n}}_i \widetilde{\lambda}_i\right]}\, .
\ee
In the next section, we will consider the Cardy limit ($|L \Omega| \ll 1$), for which the sums over $\mathfrak{n}_i$ and $\widetilde{\mathfrak{n}}_i$ in $Z_{CS}$ can be neglected. Hence, in the Cardy limit we can approximate $Z_{CS}$ by the first exponential factor of \eqref{eq:Z_CS-final}.

%%%%%%%%%%%%%%%%%%%%%%%%%%%%%%%%%%%%%%%%%%%
%%%%%%%%%%%%%%%%%%%%%%%%%%%%%%%%%%%%%%%%%%%
\section{Large-$N$ Matrix Model}\label{sec:LargeN}
%%%%%%%%%%%%%%%%%%%%%%%%%%%%%%%%%%%%%%%%%%%
%%%%%%%%%%%%%%%%%%%%%%%%%%%%%%%%%%%%%%%%%%%

Unlike the 4d case discussed in \cite{Cabo-Bizet:2018ehj}, the conformal anomaly is absent in 3d. Hence, in order to compare with the results from the gravity side, we have to perform a large-$N$ analysis of the matrix model obtained by localizing the ABJM theory, similar to the procedure in Refs.~\cite{ABJMlargeN, Benini:2015eyy}.

In order to simplify the matrix model from localization and extract the leading order contribution to the AdS$_4$ black hole entropy, we will consider two different limits in sequence. First, we will consider the Cardy limit given by $|L \Omega| \ll 1$. After that we will take the standard large-$N$ limit.

In the following we focus on the ABJM theory. From the localization result of the partition function \eqref{eq:ABJMmatrixModel} on the background discussed in Sec.~\ref{sec:BgdFields}, we obtain the free energy
\begin{align}
  F & = - \textrm{log}\, Z\nonumber\\
  {} & = \frac{k \pi}{2 L \Omega} \sum_j (\lambda_j^2 - \tilde{\lambda}_j^2) + \mathcal{O} (\Omega) \nonumber\\
  {} & \quad + 2\, \textrm{log}\, N! + 2 N\, \textrm{log}\, 2 \pi + F_c + F_v\, ,\label{eq:FreeEnergy}
\end{align}
where
\begin{align}
  F_c & \equiv - \sum_{i, j} \sum_{I=1}^4 \textrm{log} \Bigg[\frac{\left(e^{2 \pi i (\gamma - \delta_I^c)};\, e^{2 \pi i \beta} \right)_\infty}{\left(e^{2 \pi i (\gamma + \delta_I^c)};\, e^{2 \pi i \beta} \right)_\infty} \Bigg] \, ,\\
  F_v & \equiv - \sum_{i, j} \sum_{K=1}^2 \textrm{log} \Bigg[\frac{\left(e^{2 \pi i (\gamma - \delta_K^c)};\, e^{2 \pi i \beta} \right)_\infty}{\left(e^{2 \pi i (\gamma + \delta_K^c)};\, e^{2 \pi i \beta} \right)_\infty} \Bigg] \, .
\end{align}

We can  compute the leading contribution to the free energy $F$ at large $N$ analytically using the approach introduced in \cite{ABJMlargeN, Benini:2015eyy}. First, let us consider $F_c$. We apply the definitions \eqref{eq:DefParam-1} - \eqref{eq:DefParam-2} and \eqref{eq:DefDelta-1} - \eqref{eq:DefDelta-4} to rewrite it as
\begin{align}
  F_c & = - \sum_{i, j} \textrm{log} \left(e^{\frac{\lambda_i - \tilde{\lambda}_j}{2 \pi} - 2 L \Phi - \frac{i}{2} \varphi_1};\, e^{- L \Omega} \right)_\infty + \sum_{i, j} \textrm{log} \left(e^{- \frac{\lambda_i - \tilde{\lambda}_j}{2 \pi} - 6 L \Phi + \frac{i}{2} \varphi_1};\, e^{- L \Omega} \right)_\infty \nonumber\\
  {} & \quad - \sum_{i, j} \textrm{log} \left(e^{\frac{\lambda_i - \tilde{\lambda}_j}{2 \pi} - 2 L \Phi + \frac{i}{2} \varphi_1};\, e^{- L \Omega} \right)_\infty + \sum_{i, j} \textrm{log} \left(e^{- \frac{\lambda_i - \tilde{\lambda}_j}{2 \pi} - 6 L \Phi - \frac{i}{2} \varphi_1};\, e^{- L \Omega} \right)_\infty \nonumber\\
  {} & \quad - \sum_{i, j} \textrm{log} \left(e^{- \frac{\lambda_i - \tilde{\lambda}_j}{2 \pi} - 2 L \Phi - \frac{i}{2} \varphi_2};\, e^{- L \Omega} \right)_\infty + \sum_{i, j} \textrm{log} \left(e^{\frac{\lambda_i - \tilde{\lambda}_j}{2 \pi} - 6 L \Phi + \frac{i}{2} \varphi_2};\, e^{- L \Omega} \right)_\infty \nonumber\\
  {} & \quad - \sum_{i, j} \textrm{log} \left(e^{- \frac{\lambda_i - \tilde{\lambda}_j}{2 \pi} - 2 L \Phi + \frac{i}{2} \varphi_2};\, e^{- L \Omega} \right)_\infty + \sum_{i, j} \textrm{log} \left(e^{\frac{\lambda_i - \tilde{\lambda}_j}{2 \pi} - 6 L \Phi - \frac{i}{2} \varphi_2};\, e^{- L \Omega} \right)_\infty\, .\label{eq:Fc-old}
\end{align}
Due to the constraint \eqref{eq:constraint}, in the Cardy limit ($|L \Omega| \ll 1$) there is $e^{8 L \Phi} \approx e^{2 \pi i} = 1$. Hence, the exponents in \eqref{eq:Fc-old} can be shifted by multiples of $8 L \Phi$ without changing the result at the leading order, and at the leading order we can rewrite \eqref{eq:Fc-old} as
\begin{align}
  F_c & \simeq - \sum_{i, j} \textrm{log} \left(e^{\frac{\lambda_i - \tilde{\lambda}_j}{2 \pi} - 2 L \Phi - \frac{i}{2} \varphi_1};\, e^{- L \Omega} \right)_\infty + \sum_{i, j} \textrm{log} \left(e^{- \frac{\lambda_i - \tilde{\lambda}_j}{2 \pi} + 2 L \Phi + \frac{i}{2} \varphi_1};\, e^{- L \Omega} \right)_\infty \nonumber\\
  {} & \quad - \sum_{i, j} \textrm{log} \left(e^{\frac{\lambda_i - \tilde{\lambda}_j}{2 \pi} - 2 L \Phi + \frac{i}{2} \varphi_1};\, e^{- L \Omega} \right)_\infty + \sum_{i, j} \textrm{log} \left(e^{- \frac{\lambda_i - \tilde{\lambda}_j}{2 \pi} + 2 L \Phi - \frac{i}{2} \varphi_1};\, e^{- L \Omega} \right)_\infty \nonumber\\
  {} & \quad - \sum_{i, j} \textrm{log} \left(e^{- \frac{\lambda_i - \tilde{\lambda}_j}{2 \pi} - 2 L \Phi - \frac{i}{2} \varphi_2};\, e^{- L \Omega} \right)_\infty + \sum_{i, j} \textrm{log} \left(e^{\frac{\lambda_i - \tilde{\lambda}_j}{2 \pi} + 2 L \Phi + \frac{i}{2} \varphi_2};\, e^{- L \Omega} \right)_\infty \nonumber\\
  {} & \quad - \sum_{i, j} \textrm{log} \left(e^{- \frac{\lambda_i - \tilde{\lambda}_j}{2 \pi} - 2 L \Phi + \frac{i}{2} \varphi_2};\, e^{- L \Omega} \right)_\infty + \sum_{i, j} \textrm{log} \left(e^{\frac{\lambda_i - \tilde{\lambda}_j}{2 \pi} + 2 L \Phi - \frac{i}{2} \varphi_2};\, e^{- L \Omega} \right)_\infty\, .\label{eq:Fc}
\end{align}

To obtain the expression corresponding to \eqref{eq:Fc} in the continuum limit, we adopt the redefinitions
\be
  \lambda_j = \frac{N^\alpha t_j - i v_j}{\pi}\, ,\quad \tilde{\lambda}_j = \frac{N^\alpha t_j - i \tilde{v}_j}{\pi}\, ,\quad \delta v \equiv \tilde{v} - v\, ,
\ee
and the replacement
\be
  \sum_{i=1}^N \quad \Rightarrow\quad \frac{N}{2 \pi} \int dt\, \rho(t)\, .
\ee
Moreover, we apply the expansion of the $q$-Pochhammer symbol near $q=1$ \cite{Zagier}:
\be
  \textrm{log} (z,\, e^{\hbar})_\infty = \frac{1}{\hbar} \, \textrm{Li}_2 (z) + \mathcal{O} (1)
\ee
to obtain the leading contributions in the Cardy limit ($|L \Omega| \ll 1$). Consequently, in the Cardy limt the leading order of \eqref{eq:Fc} has the expression:
\begin{align}
  F_c & \simeq \frac{1}{\Omega L} \sum_{i, j} \textrm{Li}_2 \left(e^{\frac{\lambda_i - \tilde{\lambda}_j}{2 \pi} - 2 L \Phi - \frac{i}{2} \varphi_1} \right) - \frac{1}{\Omega L} \sum_{i, j} \textrm{Li}_2 \left(e^{- \frac{\lambda_i - \tilde{\lambda}_j}{2 \pi} + 2 L \Phi + \frac{i}{2} \varphi_1} \right) \nonumber\\
  {} & \quad + \frac{1}{\Omega L} \sum_{i, j} \textrm{Li}_2 \left(e^{\frac{\lambda_i - \tilde{\lambda}_j}{2 \pi} - 2 L \Phi + \frac{i}{2} \varphi_1} \right) - \frac{1}{\Omega L} \sum_{i, j} \textrm{Li}_2 \left(e^{- \frac{\lambda_i - \tilde{\lambda}_j}{2 \pi} + 2 L \Phi - \frac{i}{2} \varphi_1} \right) \nonumber\\
  {} & \quad + \frac{1}{\Omega L} \sum_{i, j} \textrm{Li}_2 \left(e^{- \frac{\lambda_i - \tilde{\lambda}_j}{2 \pi} - 2 L \Phi - \frac{i}{2} \varphi_2} \right) - \frac{1}{\Omega L} \sum_{i, j} \textrm{Li}_2 \left(e^{\frac{\lambda_i - \tilde{\lambda}_j}{2 \pi} + 2 L \Phi + \frac{i}{2} \varphi_2} \right) \nonumber\\
  {} & \quad + \frac{1}{\Omega L} \sum_{i, j} \textrm{Li}_2 \left(e^{- \frac{\lambda_i - \tilde{\lambda}_j}{2 \pi} - 2 L \Phi + \frac{i}{2} \varphi_2} \right) - \frac{1}{\Omega L} \sum_{i, j} \textrm{Li}_2 \left(e^{\frac{\lambda_i - \tilde{\lambda}_j}{2 \pi} + 2 L \Phi - \frac{i}{2} \varphi_2} \right)\, .\label{eq:FcCardylimit}
\end{align}
Let us call the 8 terms in the expression \eqref{eq:Fc} as $F_c^{(1)}$ - $F_c^{(8)}$ respectively.

Recall that the function $\textrm{Li}_2 (e^{i u})$ has the following property:
\begin{align}
\begin{split}\label{eq:property}
  \textrm{Li}_2 (e^{i u}) & = - \textrm{Li}_2 (e^{- i u}) + \frac{u^2}{2} - \pi u + \frac{\pi^2}{3}\, ,\quad\quad \textrm{for $0 < \textrm{Re} (u) < 2 \pi$}\, ;\\
  \textrm{Li}_2 (e^{i u}) & = - \textrm{Li}_2 (e^{- i u}) + \frac{u^2}{2} + \pi u + \frac{\pi^2}{3}\, ,\quad\quad \textrm{for $- 2 \pi < \textrm{Re} (u) < 0$}\, .
\end{split}
\end{align}
Using this property and assuming that
\begin{align}
\begin{split}\label{eq:assumption}
  \textrm{Re} \left(\frac{i (\lambda_i - \widetilde{\lambda}_j)}{2 \pi} - 2 i L \Phi \pm \frac{1}{2} \varphi_1 \right) & < 0\, ,\\
  \textrm{Re} \left(\frac{i (\lambda_i - \widetilde{\lambda}_j)}{2 \pi} + 2 i L \Phi \pm \frac{1}{2} \varphi_2 \right) & > 0\, ,
\end{split}
\end{align}
we can rewrite some terms in \eqref{eq:FcCardylimit} as follows:
\begin{align}
  F_c^{(2)} & = \frac{1}{\Omega L} \sum_{i, j} \Bigg[\textrm{Li}_2 \left(e^{\frac{\lambda_i - \widetilde{\lambda}_j}{2 \pi} - 2 L \Phi - \frac{i}{2} \varphi_1} \right) - \frac{1}{2} \left(\frac{i (\lambda_i - \widetilde{\lambda}_j)}{2 \pi} - 2 i L \Phi + \frac{1}{2} \varphi_1 \right)^2 \nonumber\\
  {} & \qquad\qquad - \pi \left(\frac{i (\lambda_i - \widetilde{\lambda}_j)}{2 \pi} - 2 i L \Phi + \frac{1}{2} \varphi_1 \right) - \frac{\pi^2}{3} \Bigg]\, ,
\end{align}
\begin{align}
  F_c^{(4)} & = \frac{1}{\Omega L} \sum_{i, j} \Bigg[\textrm{Li}_2 \left(e^{\frac{\lambda_i - \widetilde{\lambda}_j}{2 \pi} - 2 L \Phi + \frac{i}{2} \varphi_1} \right) - \frac{1}{2} \left(\frac{i (\lambda_i - \widetilde{\lambda}_j)}{2 \pi} - 2 i L \Phi - \frac{1}{2} \varphi_1 \right)^2 \nonumber\\
  {} & \qquad\qquad - \pi \left(\frac{i (\lambda_i - \widetilde{\lambda}_j)}{2 \pi} - 2 i L \Phi - \frac{1}{2} \varphi_1 \right) - \frac{\pi^2}{3} \Bigg]\, ,
\end{align}
\begin{align}
  F_c^{(5)} & = - \frac{1}{\Omega L} \sum_{i, j} \Bigg[\textrm{Li}_2 \left(e^{\frac{\lambda_i - \widetilde{\lambda}_j}{2 \pi} + 2 L \Phi + \frac{i}{2} \varphi_1} \right) - \frac{1}{2} \left(\frac{i (\lambda_i - \widetilde{\lambda}_j)}{2 \pi} + 2 i L \Phi - \frac{1}{2} \varphi_1 \right)^2 \nonumber\\
  {} & \qquad\qquad + \pi \left(\frac{i (\lambda_i - \widetilde{\lambda}_j)}{2 \pi} + 2 i L \Phi - \frac{1}{2} \varphi_1 \right) - \frac{\pi^2}{3} \Bigg]\, ,
\end{align}
\begin{align}
  F_c^{(7)} & = - \frac{1}{\Omega L} \sum_{i, j} \Bigg[\textrm{Li}_2 \left(e^{\frac{\lambda_i - \widetilde{\lambda}_j}{2 \pi} + 2 L \Phi - \frac{i}{2} \varphi_1} \right) - \frac{1}{2} \left(\frac{i (\lambda_i - \widetilde{\lambda}_j)}{2 \pi} + 2 i L \Phi + \frac{1}{2} \varphi_1 \right)^2 \nonumber\\
  {} & \qquad\qquad + \pi \left(\frac{i (\lambda_i - \widetilde{\lambda}_j)}{2 \pi} + 2 i L \Phi + \frac{1}{2} \varphi_1 \right) - \frac{\pi^2}{3} \Bigg]\, .
\end{align}
Consequently, \eqref{eq:FcCardylimit} becomes
\begin{align}
  F_c & \simeq \frac{2}{\Omega L} \sum_{i, j} \textrm{Li}_2 \left(e^{\frac{\lambda_i - \tilde{\lambda}_j}{2 \pi} - 2 L \Phi - \frac{i}{2} \varphi_1} \right) + \frac{2}{\Omega L} \sum_{i, j} \textrm{Li}_2 \left(e^{\frac{\lambda_i - \tilde{\lambda}_j}{2 \pi} - 2 L \Phi + \frac{i}{2} \varphi_1} \right) \nonumber\\
  {} & \quad - \frac{2}{\Omega L} \sum_{i, j} \textrm{Li}_2 \left(e^{\frac{\lambda_i - \tilde{\lambda}_j}{2 \pi} + 2 L \Phi + \frac{i}{2} \varphi_2} \right) - \frac{2}{\Omega L} \sum_{i, j} \textrm{Li}_2 \left(e^{\frac{\lambda_i - \tilde{\lambda}_j}{2 \pi} + 2 L \Phi - \frac{i}{2} \varphi_2} \right) \nonumber\\
  {} & \quad - \frac{1}{\Omega L} (4 i \pi + 8 \Phi L) \sum_{i, j} \frac{\lambda_i - \widetilde{\lambda}_j}{2 \pi} - \frac{1}{4 \Omega L} (\varphi_1^2 - \varphi_2^2) \left(\sum_{i, j} 1\right)\, .\label{eq:FcInterm}
\end{align}
Let us call the first 4 terms $F_c^{(I)}$ - $F_c^{(IV)}$ respectively, for which we carefully distinguish $i = j$, $i > j$ and $i < j$. After applying the property \eqref{eq:property} and the assumption \eqref{eq:assumption} several times, we can express $F_c^{(I)}$ - $F_c^{(IV)}$ as follows:
\begin{align}
  F_c^{(I)} & = \frac{2}{L \Omega} \frac{N}{2 \pi} \int dt\, \rho(t)\, \textrm{Li}_2 \left(e^{\frac{i}{2 \pi^2} \delta v - 2 L \Phi - \frac{i}{2} \varphi_1} \right) \nonumber\\
  {} & \quad + \frac{2 i}{L \Omega} 2 \pi \frac{N^{2 - \alpha}}{4 \pi} \int dt\, \rho^2 (t) \, g_+ \left(\frac{1}{2 \pi^2}\, \delta v(t) + 2 i L \Phi - \frac{1}{2} \varphi_1 \right) \nonumber\\
  {} & \quad - \frac{2}{L \Omega} \sum_{i > j} \Bigg[\frac{1}{2} \left(\frac{\lambda_i - \tilde{\lambda}_j}{2 \pi} - 2 L \Phi - \frac{i}{2} \varphi_1 \right)^2 - i \pi \left(\frac{\lambda_i - \tilde{\lambda}_j}{2 \pi} - 2 L \Phi - \frac{i}{2} \varphi_1 \right) - \frac{\pi^2}{3} \Bigg] \nonumber\\
  {} & \quad + \mathcal{O} (\Omega^0)\, ,\label{eq:FcI}
\end{align}
\begin{align}
  F_c^{(II)} & = \frac{2}{L \Omega} \frac{N}{2 \pi} \int dt\, \rho(t)\, \textrm{Li}_2 \left(e^{\frac{i}{2 \pi^2} \delta v - 2 L \Phi + \frac{i}{2} \varphi_1} \right) \nonumber\\
  {} & \quad + \frac{2 i}{L \Omega} 2 \pi \frac{N^{2 - \alpha}}{4 \pi} \int dt\, \rho^2 (t) \, g_+ \left(\frac{1}{2 \pi^2}\, \delta v(t) + 2 i L \Phi + \frac{1}{2} \varphi_1 \right) \nonumber\\
  {} & \quad - \frac{2}{L \Omega} \sum_{i > j} \Bigg[\frac{1}{2} \left(\frac{\lambda_i - \tilde{\lambda}_j}{2 \pi} - 2 L \Phi + \frac{i}{2} \varphi_1 \right)^2 - i \pi \left(\frac{\lambda_i - \tilde{\lambda}_j}{2 \pi} - 2 L \Phi + \frac{i}{2} \varphi_1 \right) - \frac{\pi^2}{3} \Bigg] \nonumber\\
  {} & \quad + \mathcal{O} (\Omega^0)\, ,\label{eq:FcII}
\end{align}
\begin{align}
  F_c^{(III)} & = - \frac{2}{L \Omega} \frac{N}{2 \pi} \int dt\, \rho(t)\, \textrm{Li}_2 \left(e^{\frac{i}{2 \pi^2} \delta v + 2 L \Phi + \frac{i}{2} \varphi_2} \right) \nonumber\\
  {} & \quad - \frac{2 i}{L \Omega} 2 \pi \frac{N^{2 - \alpha}}{4 \pi} \int dt\, \rho^2 (t) \, g_- \left(\frac{1}{2 \pi^2}\, \delta v(t) - 2 i L \Phi + \frac{1}{2} \varphi_2 \right) \nonumber\\
  {} & \quad + \frac{2}{L \Omega} \sum_{i > j} \Bigg[\frac{1}{2} \left(\frac{\lambda_i - \tilde{\lambda}_j}{2 \pi} + 2 L \Phi + \frac{i}{2} \varphi_2 \right)^2 + i \pi \left(\frac{\lambda_i - \tilde{\lambda}_j}{2 \pi} + 2 L \Phi + \frac{i}{2} \varphi_2 \right) - \frac{\pi^2}{3} \Bigg] \nonumber\\
  {} & \quad + \mathcal{O} (\Omega^0)\, ,\label{eq:FcIII}
\end{align}
\begin{align}
  F_c^{(IV)} & = - \frac{2}{L \Omega} \frac{N}{2 \pi} \int dt\, \rho(t)\, \textrm{Li}_2 \left(e^{\frac{i}{2 \pi^2} \delta v + 2 L \Phi - \frac{i}{2} \varphi_2} \right) \nonumber\\
  {} & \quad - \frac{2 i}{L \Omega} 2 \pi \frac{N^{2 - \alpha}}{4 \pi} \int dt\, \rho^2 (t) \, g_- \left(\frac{1}{2 \pi^2}\, \delta v(t) - 2 i L \Phi - \frac{1}{2} \varphi_2 \right) \nonumber\\
  {} & \quad + \frac{2}{L \Omega} \sum_{i > j} \Bigg[\frac{1}{2} \left(\frac{\lambda_i - \tilde{\lambda}_j}{2 \pi} + 2 L \Phi - \frac{i}{2} \varphi_2 \right)^2 + i \pi \left(\frac{\lambda_i - \tilde{\lambda}_j}{2 \pi} + 2 L \Phi - \frac{i}{2} \varphi_2 \right) - \frac{\pi^2}{3} \Bigg] \nonumber\\
  {} & \quad + \mathcal{O} (\Omega^0)\, ,\label{eq:FcIV}
\end{align}
where
\be
  g_- (x) \equiv \frac{x^3}{6} + \frac{\pi}{2} x^2 + \frac{\pi^2}{3} x\, ,\quad g_+ (x) \equiv \frac{x^3}{6} - \frac{\pi}{2} x^2 + \frac{\pi^2}{3} x\, .
\ee
The sum of the third lines of \eqref{eq:FcI} - \eqref{eq:FcIV} is
\be
  \frac{2}{\Omega L} (4 i \pi + 8 \Phi L) \sum_{i > j} \frac{\lambda_i - \widetilde{\lambda}_j}{2 \pi} + \frac{1}{2 \Omega L} (\varphi_1^2 - \varphi_2^2) \left(\sum_{i > j} 1\right)\, ,
\ee
which cancels the third line of \eqref{eq:FcInterm} up to a constant of order $\mathcal{O} (N)$. Therefore, the final expression of the leading terms of $F_c$ in the Cardy limit is given by the sum of the first and second lines of \eqref{eq:FcI} - \eqref{eq:FcIV}.

For the vector multiplet we obtain
\begin{align}
  F_v & = - \sum_{i\neq j} \textrm{log} \left(e^{\frac{\lambda_i - \lambda_j}{2 \pi} - 8 L \Phi};\, e^{- L \Omega} \right)_\infty + \sum_{i\neq j} \textrm{log} \left(e^{- \frac{\lambda_i - \lambda_j}{2 \pi}};\, e^{- L \Omega} \right)_\infty \nonumber\\
  {} & \quad - \sum_{i\neq j} \textrm{log} \left(e^{- \frac{\tilde{\lambda}_i - \tilde{\lambda}_j}{2 \pi} - 8 L \Phi};\, e^{- L \Omega} \right)_\infty + \sum_{i\neq j} \textrm{log} \left(e^{\frac{\tilde{\lambda}_i - \tilde{\lambda}_j}{2 \pi}};\, e^{- L \Omega} \right)_\infty\, .\label{eq:Fv}
\end{align}
At the leading order, each term of $F_v$ can be expressed as
\begin{align}
  F_v^{(1)} & \simeq \frac{1}{L \Omega} \sum_{i \neq j} \textrm{Li}_2 \left(e^{\frac{\lambda_i - \lambda_j}{2 \pi} - 8 L \Phi} \right) \, ,\\
  F_v^{(2)} & \simeq - \frac{1}{L \Omega} \sum_{i \neq j} \textrm{Li}_2 \left(e^{- \frac{\lambda_i - \lambda_j}{2 \pi}} \right) \, ,\\
  F_v^{(3)} & \simeq \frac{1}{L \Omega} \sum_{i \neq j} \textrm{Li}_2 \left(e^{\frac{\tilde{\lambda}_i - \tilde{\lambda}_j}{2 \pi} - 8 L \Phi} \right) \, ,\\
  F_v^{(4)} & \simeq - \frac{1}{L \Omega} \sum_{i \neq j} \textrm{Li}_2 \left(e^{- \frac{\tilde{\lambda}_i - \tilde{\lambda}_j}{2 \pi}} \right) \, .
\end{align}
They cancel each other at the leading order of the Cardy limit, because in this limit $e^{- 8 L \Phi} \approx 1$.

Putting everything together and neglecting some subleading terms, we obtain the free energy in both the large-$N$ and the Cardy limit:
\begin{align}
  F & \simeq N^{1 + \alpha} \frac{i k}{L \Omega} \int dt\, t\, \rho(t)\, \widetilde{\delta v} (t) \nonumber\\
  {} & \quad + N \frac{2}{L \Omega} \frac{1}{2 \pi} \int dt\, \rho(t) \Bigg[\textrm{Li}_2 \left(e^{i \left[\widetilde{\delta v} (t) + i \Delta_3 \right]} \right) + \textrm{Li}_2 \left(e^{i \left[\widetilde{\delta v} (t) + i \Delta_4 \right]} \right) \nonumber\\
  {} & \qquad\qquad\qquad\qquad\qquad - \textrm{Li}_2 \left(e^{i \left[\widetilde{\delta v} (t) - i \Delta_1 \right]} \right) - \textrm{Li}_2 \left(e^{i \left[\widetilde{\delta v} (t) - i \Delta_2 \right]} \right) \Bigg] \nonumber\\
  {} & \quad + N^{2 - \alpha} \frac{4 i \pi}{L \Omega} \frac{1}{4 \pi} \int dt\, \rho^2 (t) \Bigg[ g_+ \left( \widetilde{\delta v} (t) + i \Delta_3 \right) + g_+ \left( \widetilde{\delta v} (t) + i \Delta_4 \right) \nonumber\\
  {} & \qquad\qquad\qquad\qquad\qquad\qquad - g_- \left( \widetilde{\delta v} (t) - i \Delta_1 \right) - g_- \left( \widetilde{\delta v} (t) - i \Delta_2 \right) \Bigg] \, ,\label{eq:largeN-F-1}
\end{align}
where
\be
  \widetilde{\delta v} (t) \equiv \frac{1}{2 \pi^2} \delta v (t)\, ,
\ee
\be
  \Delta_3 \equiv 2 L \Phi + \frac{i}{2} \varphi_1\, ,\quad \Delta_4 \equiv 2 L \Phi - \frac{i}{2} \varphi_1\, ,
\ee
\be
  \Delta_1 \equiv 2 L \Phi + \frac{i}{2} \varphi_2\, ,\quad \Delta_2 \equiv 2 L \Phi - \frac{i}{2} \varphi_2\, .
\ee
In order that the terms of the orders $\mathcal{O} (N^{1+\alpha})$ and $\mathcal{O} (N^{2-\alpha})$ in \eqref{eq:largeN-F-1} compete with each other, we further require that
\be
  \alpha = \frac{1}{2}\, .
\ee
Hence, we obtain the leading contribution of the free energy $F$ at the order $\mathcal{O} (\Omega^{-1})$ in both the large-$N$ and the Cardy limit:
\begin{align}
  F & \simeq N^{\frac{3}{2}} \frac{i k}{L \Omega} \int dt\, t\, \rho(t)\, \widetilde{\delta v} (t) \nonumber\\
  {} & \quad + N \frac{1}{\pi L \Omega} \int dt\, \rho(t) \Bigg[\textrm{Li}_2 \left(e^{i \left[\widetilde{\delta v} (t) + i \Delta_3 \right]} \right) + \textrm{Li}_2 \left(e^{i \left[\widetilde{\delta v} (t) + i \Delta_4 \right]} \right) \nonumber\\
  {} & \qquad\qquad\qquad\qquad\qquad - \textrm{Li}_2 \left(e^{i \left[\widetilde{\delta v} (t) - i \Delta_1 \right]} \right) - \textrm{Li}_2 \left(e^{i \left[\widetilde{\delta v} (t) - i \Delta_2 \right]} \right) \Bigg] \nonumber\\
  {} & \quad + N^{\frac{3}{2}} \frac{i}{L \Omega} \int dt\, \rho^2 (t) \Bigg[ g_+ \left( \widetilde{\delta v} (t) + i \Delta_3 \right) + g_+ \left( \widetilde{\delta v} (t) + i \Delta_4 \right) \nonumber\\
  {} & \qquad\qquad\qquad\qquad\qquad - g_- \left( \widetilde{\delta v} (t) - i \Delta_1 \right) - g_- \left( \widetilde{\delta v} (t) - i \Delta_2 \right) \Bigg] \, .\label{eq:largeN-F-2}
\end{align}
Although the second integral of \eqref{eq:largeN-F-2} is of the order $\mathcal{O} (N)$, while the first and the third integrals are of the order $\mathcal{O} (N^{\frac{3}{2}})$, the second line will contribute to the saddle point solutions. We can add a new term with the Lagrange multiplier $\mu$ that imposes the normalization condition $\int dt\, \rho(t) = 1$. Consequently,
\begin{align}
  \frac{L \Omega}{i N^{\frac{3}{2}}} F & \simeq k \int dt\, t\, \rho(t)\, \widetilde{\delta v} (t) - \mu \Big[\int dt\, \rho(t) - 1 \Big] \nonumber\\
  {} & \quad - \frac{i}{\pi N^{\frac{1}{2}}} \int dt\, \rho(t) \Bigg[\textrm{Li}_2 \left(e^{i \left[\widetilde{\delta v} (t) + i \Delta_3 \right]} \right) + \textrm{Li}_2 \left(e^{i \left[\widetilde{\delta v} (t) + i \Delta_4 \right]} \right) \nonumber\\
  {} & \qquad\qquad\qquad\qquad - \textrm{Li}_2 \left(e^{i \left[\widetilde{\delta v} (t) - i \Delta_1 \right]} \right) - \textrm{Li}_2 \left(e^{i \left[\widetilde{\delta v} (t) - i \Delta_2 \right]} \right) \Bigg] \nonumber\\
  {} & \quad + \int dt\, \rho^2 (t) \Bigg[ g_+ \left( \widetilde{\delta v} (t) + i \Delta_3 \right) + g_+ \left( \widetilde{\delta v} (t) + i \Delta_4 \right) \nonumber\\
  {} & \qquad\qquad\qquad - g_- \left( \widetilde{\delta v} (t) - i \Delta_1 \right) - g_- \left( \widetilde{\delta v} (t) - i \Delta_2 \right) \Bigg] \, .\label{eq:largeN-F-3}
\end{align}
This is precisely the Bethe potential considered in \cite{Benini:2015eyy} and the matrix model obtained from the superconformal index \cite{Choi:2019zpz}.

We can follow the same steps as in \cite{Benini:2015eyy} to analyze the saddle-point contributions to the free energy $F$. Let us first consider the case $k=1$. We derive the saddle-point equations from the large-$N$ free energy in the Cardy limit as follows:
\be
  \frac{L \Omega}{i N^{\frac{3}{2}}} \frac{\partial F}{\partial \rho} = 0\, ,\quad \frac{L \Omega}{i N^{\frac{3}{2}}} \frac{\partial F}{\partial \widetilde{\delta v}} = 0\, .\label{eq:SaddlePtEqs}
\ee
Similar to \cite{Benini:2015eyy}, we define $Y_I$ through
\begin{align}
\begin{split}
  \widetilde{\delta v} \equiv i e^{- N^\frac{1}{2} Y_3} - i \Delta_3\, , & \quad \widetilde{\delta v} \equiv i e^{- N^\frac{1}{2} Y_4} - i \Delta_4\, ,\\
  \widetilde{\delta v} \equiv i e^{- N^\frac{1}{2} Y_1} + i \Delta_1\, , & \quad \widetilde{\delta v} \equiv i e^{- N^\frac{1}{2} Y_2} + i \Delta_2\, ,
\end{split}
\end{align}
and these special values of $\widetilde{\delta v}$ are called $(\widetilde{\delta v})_*$. The saddle-point equations \eqref{eq:SaddlePtEqs} become
\be
  t\, \widetilde{\delta v} + 2 \rho \Bigg[ g_+ \left(\widetilde{\delta v} (t) + i \Delta_3 \right) + g_+ \left(\widetilde{\delta v} (t) + i \Delta_4 \right) - g_- \left(\widetilde{\delta v} (t) - i \Delta_1 \right) - g_- \left(\widetilde{\delta v} (t) - i \Delta_2 \right) \Bigg] = \mu\, ,\label{eq:SaddlePtEq-1}
\ee
when $\widetilde{\delta v} \napprox (\widetilde{\delta v})_*$:
\be
  t + \rho \Bigg[ g_+' \left(\widetilde{\delta v} (t) + i \Delta_3 \right) + g_+' \left(\widetilde{\delta v} (t) + i \Delta_4 \right) - g_-' \left(\widetilde{\delta v} (t) - i \Delta_1 \right) - g_-' \left(\widetilde{\delta v} (t) - i \Delta_2 \right) \Bigg] = 0\, ,\label{eq:SaddlePtEq-2}
\ee
when $\widetilde{\delta v} = (\widetilde{\delta v})_*$:
\begin{align}
  {} & t + \rho \Bigg[ g_+' \left(\widetilde{\delta v} (t) + i \Delta_3 \right) + g_+' \left(\widetilde{\delta v} (t) + i \Delta_4 \right) - g_-' \left(\widetilde{\delta v} (t) - i \Delta_1 \right) - g_-' \left(\widetilde{\delta v} (t) - i \Delta_2 \right) \Bigg] \nonumber\\
  = & \, - \frac{1}{\pi} (Y_3 + Y_4 - Y_1 - Y_2)\, .\label{eq:SaddlePtEq-3}
\end{align}

Let us first define
\be
  t_\ll \equiv - \frac{\mu}{i \Delta_3}\, ,\quad t_< \equiv - \frac{\mu}{i \Delta_4}\, ,\quad t_> \equiv \frac{\mu}{i \Delta_2}\, ,\quad t_\gg \equiv \frac{\mu}{i \Delta_1}\, .
\ee
The sadde-point equations \eqref{eq:SaddlePtEq-1} - \eqref{eq:SaddlePtEq-3} can be solved in different intervals:
\begin{enumerate}
\item $t \in [t_\ll,\, t_<]$:
\begin{align}
  \rho & = \frac{\mu + t\, i \Delta_3}{(i \Delta_1 + i \Delta_3) (i \Delta_2 + i \Delta_3) (i \Delta_4 - i \Delta_3)}\, ,\\
  \delta v & = - i \Delta_3\, ,\\
  Y_3 & = \frac{\pi (- t\, i \Delta_4 - \mu)}{i \Delta_4 - i \Delta_3}\, .
\end{align}

\item $t \in [t_<,\, t_>]$:
\begin{align}
  \rho & = \frac{2 \pi \mu + t (i \Delta_3\, i \Delta_4 - i \Delta_1\, i \Delta_2)}{(i \Delta_1 + i \Delta_3) (i \Delta_2 + i \Delta_3) (i \Delta_1 + i \Delta_4) (i \Delta_2 + i \Delta_4)}\, ,\\
  \delta v & = \frac{\mu (i \Delta_1\, i \Delta_2 - i \Delta_3\, i \Delta_4) + t \sum_{I < J < K} i \Delta_I\, i \Delta_J\, i \Delta_K}{2 \pi \mu + t (i \Delta_3\, i \Delta_4 - i \Delta_1\, i \Delta_2)}\, .
\end{align}

\item $t \in [t_>,\, t_\gg]$:
\begin{align}
  \rho & = \frac{\mu - t\, i \Delta_1}{(i \Delta_1 + i \Delta_3) (i \Delta_1 + i \Delta_4) (i \Delta_2 - i \Delta_1)}\, ,\\
  \delta v & = i \Delta_1\, ,\\
  Y_1 & = \frac{\pi (t\, i \Delta_2 - \mu)}{i \Delta_2 - i \Delta_1}\, .
\end{align}
\end{enumerate}
Outside these intervals the function $\rho$ vanishes, hence we only need to consider the solutions in these intervals.

By requiring that
\be
  \int dt\, \rho (t) = 1\, ,
\ee
we can fix the constant $\mu$. Applying the constraint \eqref{eq:constraint}, we obtain the expression of $\mu$ at the leading order:
\be
  \mu = \sqrt{2 \Delta_1 \Delta_2 \Delta_3 \Delta_4}\, .
\ee
Plugging all the solutions back into the large-$N$ matrix model \eqref{eq:largeN-F-2}, we obtain at the free energy at the saddle points for $k = 1$:
\be\label{eq:largeN-F-4}
  F \simeq \frac{i N^{\frac{3}{2}}}{L \Omega} \frac{2 \sqrt{2 \Delta_1 \Delta_2 \Delta_3 \Delta_4}}{3}\, .
\ee
We can define
\be
  \omega \equiv L \Omega\, ,
\ee
then the free energy \eqref{eq:largeN-F-4} at the saddle points for $k=1$ becomes
\be\label{eq:largeN-F-5}
  F \simeq \frac{2 \sqrt{2}\, i\, N^{\frac{3}{2}}}{3} \frac{\sqrt{\Delta_1 \Delta_2 \Delta_3 \Delta_4}}{\omega}\, ,
\ee
and now the constraint \eqref{eq:constraint} is
\be\label{eq:constraintNew}
  \sum_I \Delta_I - 2 \omega = 2 \pi i\quad (\textrm{mod } 4 \pi i)\, .
\ee

In order to obtain the free energy at arbitrary value of $k$, similar to \cite{Choi:2019zpz}, we can redefine $\rho = k \hat{\rho}$ and perform similar steps. The final result of the real free energy at arbitrary $k$ is
\be
\label{Eq:MainResult}
  F \simeq \frac{2 \sqrt{2}\, i\, k^{\frac{1}{2}} N^{\frac{3}{2}}}{3} \frac{\sqrt{\Delta_1 \Delta_2 \Delta_3 \Delta_4}}{\omega}\, ,
\ee
obeying the same constraint \eqref{eq:constraintNew}.

%%%%%%%%%%%%%%%%%%%%%%%%%%%%%%%%%%%%%%%%%%%
%%%%%%%%%%%%%%%%%%%%%%%%%%%%%%%%%%%%%%%%%%%
\section{Electrically Charged AdS$_4$ Black Hole Entropy}\label{sec:BHEntropyFct}
%%%%%%%%%%%%%%%%%%%%%%%%%%%%%%%%%%%%%%%%%%%
%%%%%%%%%%%%%%%%%%%%%%%%%%%%%%%%%%%%%%%%%%%

In the large-$N$ limit, the ABJM theory is dual to M-theory on AdS$_4 \times S^7 / \mathbb{Z}_k$ \cite{Aharony:2008ug}. Hence, we expect that the free energy of the ABJM theory in both the large-$N$ and the Cardy limits obtained in the previous section should correspond to the free energy of the electrically charged AdS$_4$ black hole according to the AdS/CFT correspondence. In this section we briefly review how to obtain the  entropy of the rotating electrically charged AdS$_4$ black hole  from the entropy function  \cite{Choi:2018fdc, Cassani:2019mms}. A microscopic derivation of this entropy function was obtained in  \cite{Choi:2019zpz} starting from the superconformal index; our result  in equation (\ref{Eq:MainResult}) provides an alternative microscopic description.

Before we discuss the microscopic black hole entropy, there is a conceptual issue that we would like to emphasize.
The rotating electrically charged AdS$_4$ black holes discussed in this section are solutions in the 4d Lorentzian supergravity theory, whose holography has been studied in \cite{Hristov:2013spa}. However, in order to perform the localization, we have to consider the 3d supersymmetric field theories on the Euclidean boundary manifold in Sec.~\ref{sec:BgdFields} through Sec.~\ref{sec:LargeN}. In spite of the different signatures, the partition function $Z$ and the large-$N$ free energy $F$ obtained for the Euclidean boundary manifold can still be used to study the asymptotically AdS$_4$ black holes in the Lorentzian supergravity theory, because the Killing spinors found in Sec.~\ref{sec:BgdFields} for the Euclidean boundary manifold can be Wick rotated easily to produce the Killing spinors for the Lorentzian boundary manifold. Hence, we can interpret the results of $Z$ and $F$ obtained in the previous sections as the ones for the Lorentzian signature via an analytic continuation.

We first define an entropy function via a Legendre transformation of $\textrm{log}\, Z = - F$:
\be\label{eq:EntropyFct-1}
  S (\Delta_I,\, \omega) = - \frac{2 \sqrt{2}\, i\, k^{\frac{1}{2}} N^{\frac{3}{2}}}{3} \frac{\sqrt{\Delta_1 \Delta_2 \Delta_3 \Delta_4}}{\omega} + 2 \omega J + \sum_I \Delta_I Q_I + \Lambda \left(\sum_I \Delta_I - 2 \omega - 2 \pi i \right)\, ,
\ee
where $2 \omega$ instead of $\omega$ was treated as a variable,  because $2 \omega$ appears in the constraint.
The entropy function depends on the potentials $(\Delta_I,\, \omega)$, while the electric charges $Q_I$ and the angular momentum $J$ are introduced through the Legendre transformation. By extremizing $S (\Delta_I,\, \omega)$ as a function of $(\Delta_I,\, \omega)$, we can express the potentials $(\Delta_I,\, \omega)$ in terms of $(Q_I,\, J)$. Plugging these solutions back into \eqref{eq:EntropyFct-1}, we obtain the BPS black hole entropy.

Now let us define two new variables
\be
  \widetilde{\Delta}_I \equiv \Delta_I\, ,\quad \widetilde{\omega} \equiv 2 \omega
\ee
satisfying the following constraint equivalent to \eqref{eq:constraintNew}:
\be\label{eq:constraintNew-2}
  \sum_I \widetilde{\Delta}_I - \widetilde{\omega} = 2 \pi i\quad (\textrm{mod } 4 \pi i)\, .
\ee
Using these new variables, we can express the entropy function as
\be\label{eq:EntropyFct-2}
  S (\widetilde{\Delta}_I,\, \widetilde{\omega}) = - \frac{4 \sqrt{2}\, i\, k^{\frac{1}{2}} N^{\frac{3}{2}}}{3} \frac{\sqrt{\widetilde{\Delta}_1 \widetilde{\Delta}_2 \widetilde{\Delta}_3 \widetilde{\Delta}_4}}{\widetilde{\omega}} + \widetilde{\omega} J + \sum_I \widetilde{\Delta}_I Q_I + \Lambda \left(\sum_I \widetilde{\Delta}_I - \widetilde{\omega} - 2 \pi i \right)\, .
\ee
We extremize the entropy function \eqref{eq:EntropyFct-2} by solving the equations:
\be
  \frac{\partial S}{\partial \widetilde{\Delta}_I} = 0\, ,\quad \frac{\partial S}{\partial \widetilde{\omega}} = 0\, ,
\ee
more explicitly,
\begin{align}
  Q_I + \Lambda & = \frac{4 \sqrt{2}\, i\, k^{\frac{1}{2}} N^{\frac{3}{2}}}{3} \frac{\sqrt{\widetilde{\Delta}_1 \widetilde{\Delta}_2 \widetilde{\Delta}_3 \widetilde{\Delta}_4}}{2 \widetilde{\Delta}_I \widetilde{\omega}}\, ,\label{eq:ExtEq-1}\\
  J - \Lambda & = - \frac{4 \sqrt{2}\, i\, k^{\frac{1}{2}} N^{\frac{3}{2}}}{3} \frac{\sqrt{\widetilde{\Delta}_1 \widetilde{\Delta}_2 \widetilde{\Delta}_3 \widetilde{\Delta}_4}}{\widetilde{\omega}^2}\, .\label{eq:ExtEq-2}
\end{align}
Substituting these equations into the entropy function \eqref{eq:EntropyFct-2}, we obtain
\be\label{eq:S_Interm}
  S = - 2 \pi i \Lambda\, .
\ee
Moreover, we can combine the equations \eqref{eq:ExtEq-1} and \eqref{eq:ExtEq-2} into one equation:
\begin{align}\label{eq:ExtEq-3}
  {} & \, Q_1 Q_2 Q_3 Q_4 + \Lambda \left(\sum_{I < J < K} Q_I Q_J Q_K \right) + \Lambda^2 \left(\sum_{I < J} Q_I Q_J \right) + \Lambda^3 \left(\sum_I Q_I \right) + \Lambda^4 \nonumber\\
  = & \, - \frac{2}{9} k N^3 (\Lambda^2 - 2 \Lambda J + J^2)\, .
\end{align}
In order for the entropy $S$ to be real-valued, $\Lambda$ should be purely imaginary according to \eqref{eq:S_Interm}. We can separate the real and the imaginary parts of \eqref{eq:ExtEq-3} to obtain two independent real-valued equations:
\begin{align}
  \Lambda^4 + \Lambda^2 \left(\sum_{I < J} Q_I Q_J \right) + Q_1 Q_2 Q_3 Q_4 & = - \frac{2}{9} k N^3 \Lambda^2 - \frac{2}{9} k N^3 J^2\, ,\label{eq:ExtEq-4}\\
  \Lambda^3 \left(\sum_I Q_I \right) + \Lambda \left(\sum_{I < J < K} Q_I Q_J Q_K \right) & = \frac{4}{9} k N^3 J \Lambda\, .\label{eq:ExtEq-5}
\end{align}
We then solve \eqref{eq:ExtEq-5} to obtain $\Lambda$ and consequently the black hole entropy $S_{BH}$. By substituting the solution of $\Lambda$ into \eqref{eq:ExtEq-4}, we will obtain a constraint on $J$ and $Q_I$'s.

To simplify the discussions, we consider a degenerate case $Q_1 = Q_3$, $Q_2 = Q_4$, and the corresponding solution to $\Lambda$ and the constraint on $J$ and $Q_I$'s are
\be\label{eq:SolLambda}
  \Lambda = \pm \frac{i}{3} \sqrt{\frac{9 Q_1 Q_2 (Q_1 + Q_2) - 2 k J N^3}{Q_1 + Q_2}}\, ,
\ee
\be\label{eq:BHconstr}
  2 k J^2 N^3 + 2 k J N^3 (Q_1 + Q_2) - 9 Q_1 Q_2 (Q_1 + Q_2)^2 = 0\, .
\ee
We take the minus sign in the solution to $\Lambda$, in order for the black hole entropy to be non-negative, which has the value
\be\label{eq:S_BH-1}
  S_{BH} = \frac{2 \pi}{3} \sqrt{\frac{9 Q_1 Q_2 (Q_1 + Q_2) - 2 k J N^3}{Q_1 + Q_2}}\, .
\ee

In fact, the black hole entropy $S_{BH}$ and the angular momentum $J$ can be expressed in an alternative way. From \eqref{eq:BHconstr} we obtain
\be
  9 Q_1 Q_2 (Q_1 + Q_2) - 2 k J N^3 = \frac{2 k J^2 N^3}{Q_1 + Q_2}\, .
\ee
By plugging it into \eqref{eq:S_BH-1}, we have another expression of the black hole entropy
\be\label{eq:S_BH-2}
  S_{BH} = \frac{2 \sqrt{2} \pi k^{\frac{1}{2}} N^{\frac{3}{2}}}{3} \frac{J}{Q_1 + Q_2}\, .
\ee
We can also view \eqref{eq:BHconstr} as a quadratic equation for $J$. By solving it, we obtain the expression of $J$
\be\label{eq:Sol_J}
  J = \frac{1}{2} (Q_1 + Q_2) \left(- 1 \pm \sqrt{1 + \frac{18 Q_1 Q_2}{k N^3}} \right)\, .
\ee
If we identify some parameters with the ones on the gravity side in the following way:
\be
  \frac{1}{G} = \frac{2 \sqrt{2}}{3} g^2 k^{\frac{1}{2}} N^{\frac{3}{2}}\, ,\quad Q_{BH} = \frac{g}{2} Q\, ,\quad J_{BH} = J\, ,
\ee
we can rewrite the black hole entropy \eqref{eq:S_BH-2} and the angular momentum \eqref{eq:Sol_J} as
\begin{align}
  S_{BH} & = \frac{\pi}{g^2 G} \frac{J_{BH}}{ \left( \frac{2}{g} Q_{BH, 1} + \frac{2}{g} Q_{BH, 2} \right)}\, ,\label{eq:S_BH-3}\\
  J_{BH} & = \frac{1}{2} \left(\frac{2}{g} Q_{BH, 1} + \frac{2}{g} Q_{BH, 2} \right) \left(- 1 + \sqrt{1 + 16 g^4 G^2 \frac{2 Q_{BH, 1}}{g} \frac{2 Q_{BH, 2}}{g}} \right)\, ,\label{eq:Sol_J-2}
\end{align}
where we assume that $J_{BH} > 0$. The expressions \eqref{eq:S_BH-3} and \eqref{eq:Sol_J-2} match exactly the results on the gravity side \cite{Chong:2004na, Cvetic:2005zi, Choi:2018fdc}.  Recently, more general AdS$_4$ solutions with four generic electric charges $Q_I$'s have been constructed in \cite{Hristov:2019mqp}.

%%%%%%%%%%%%%%%%%%%%%%%%%%%%%%%%%%%%%%%%%%%
%%%%%%%%%%%%%%%%%%%%%%%%%%%%%%%%%%%%%%%%%%%
\section{Discussion}\label{sec:Discussion}
%%%%%%%%%%%%%%%%%%%%%%%%%%%%%%%%%%%%%%%%%%%
%%%%%%%%%%%%%%%%%%%%%%%%%%%%%%%%%%%%%%%%%%%

In this manuscript  we compute the partition function of 3d $\mathcal{N}=2$ gauge theories on curved spaces coinciding with the asymptotics of rotating electically charged supersymmetric AdS$_4$ black holes. We have turned on some complex background fields to preserve a pair of independent supercharges with anti-periodic boundary conditions along $S^1$, matching the amount of supersymmetry of AdS$_4$ black holes in 4d $\mathcal{N}=2$ gauged supergravity. We then applied supersymmetric localization to compute the corresponding  partition functions. As a special example, we consider the ABJM theory on this background, whose partition function in the large-$N$ and a Cardy-like  limits successfully produces the entropy function and the corresponding black hole entropy of a class of rotating electrically charged BPS AdS$_4$ black holes. Our approach complements the previous microscopic explanation of the black hole entropy based on the superconformal index of the ABJM  field theory \cite{Choi:2019zpz}.

There are various  conceptual issues that deserve further investigation. For instance, although the starting points for free energy computations are not the same, different approaches all lead effectively to the same matrix model. Originally the matrix model was studied in \cite{Benini:2015eyy} as the result of the large$-N$ limit of the topologically twisted index in a successful microscopic description of  magnetically charged BPS AdS$_4$ black holes.  The very same effective matrix model resurfaced in the study of the superconformal index of ABJM theory  \cite{Choi:2019zpz} as a microscopic description of rotating, electrically charged BPS AdS$_4$ black holes. Finally, this effective matrix model shows in our work which is based on supersymmetric localization. It would be quite  interesting  to gain a better understanding of this coincidence at a more fundamental level. There seem to be universal relations among these various partitions functions which might even reach the sphere partition function $Z_{S^3}$.   Related to this observation, some other universal relations among conformal anomaly coefficients and partition functions in different dimensions have been previously found in \cite{Benini:2015bwz, Bobev:2017uzs} and more recently in \cite{Bobev:2019zmz}. At the technical level, adding magnetic charges or electric charges and angular momentum simply  changes an overall factor in the same matrix model.   An attack on this question can probably be mounted following direct computations of indices and partition functions such as \cite{Hosseini:2016tor, Hosseini:2016ume, Jain:2019lqb, Amariti:2019pky, Jain:2019euv}. A rigorous understanding of these universal relations can possibly be achieved from the insightful work \cite{Closset:2017zgf} and the formula therein $Z_{\mathcal{M}_{q, p}}$ with an additional refinement.

We can generalize our results on BPS AdS$_4$ black hole entropy in various ways. Since the supersymmetric localization provides the exact results, besides the leading order discussed in this paper, we may extract the subleading 1-loop corrections to the black hole entropy from the exact partition function of the boundary ABJM theory. As various previous works has demonstrated  \cite{Gupta:2014hxa, Murthy:2015yfa, Nian:2017hac, Jeon:2017aif, Liu:2017vll, Liu:2017vbl,Benini:2019dyp, Hristov:2018lod, Liu:2018bac, Hristov:2019xku} such an analysis can provide more precise tests of the  AdS/CFT correspondence. Going beyond the Cardy-like limit, however, seems more formidable. The absence of a Bethe Ansatz description for the superconformal index seems to block a route that was successfully taken in the 4d context \cite{Lezcano:2019pae, Lanir:2019abx}. It would be interesting to consider near-BPS configurations by turning on temperature or slightly violating the BPS constraint \eqref{eq:constraintNew}, in the same spirit as \cite{Larsen:2019oll}. The same technique, in principle, can be applied to more general black holes, e.g. AdS$_4$ dyonic black holes, or AdS black holes in other dimensions. Moreover, we should be able to introduce M5-branes and use the new techniques to study a system of M2-branes suspended between parallel M5-branes described by ABJM theory with some appropriate boundary conditions \cite{Haghighat:2013gba, Hosomichi:2014rqa}.

Beyond  the exciting applications to black hole entropy counting, the supersymmetric localization applied to the complex background itself is very important progress. Similar to the 4d case \cite{Cabo-Bizet:2018ehj},  these new field theoretic observables are among the first that can be constructed relying on  spinors with anti-periodic boundary condition along $S^1$ without completely breaking supersymmetry. Supersymmetric localization is thus providing a new class of, in principle, exact nonperturbative observables.  It is a well-defined and interesting field theoretic question to study these observables in supersymmetric 3d field theories in general and, eventually, use them to connect with other important methods such as the conformal bootstrap.

%%%%%%%%%%%%%%%%%%%%%%%%%%%%%%%%%%%%%%%%%%%
%%%%%%%%%%%%%%%%%%%%%%%%%%%%%%%%%%%%%%%%%%%
\section*{Acknowledgements}
%%%%%%%%%%%%%%%%%%%%%%%%%%%%%%%%%%%%%%%%%%%
%%%%%%%%%%%%%%%%%%%%%%%%%%%%%%%%%%%%%%%%%%%

We would like to than Francesco Benini, Nikolay Bobev, Sunjin Choi, Dongmin Gang, Alba Grassi, Seyed Morteza Hosseini, Kiril Hristov, Chiung Hwang, Dharmesh Jain, Joonho Kim, Seok Kim, Finn Larsen, Alfredo Gonz\'alez Lezcano, Shiraz Minwalla, June Nahmgoong, Elli Pomoni, Wei Song, Qiang Wen, Jianfei Xu, Itamar Yaakov and Yang Zhou for many helpful discussions. This  work was supported in part by the U.S. Department of Energy under grant DE-SC0007859. J.N was also supported by a Van Loo Postdoctoral Fellowship and he  would like to thank the Southeast University and Keio University for hospitality during the final stages of this work.

\appendix
%%%%%%%%%%%%%%%%%%%%%%%%%%%%%%%%%%%%%%%%%%%
%%%%%%%%%%%%%%%%%%%%%%%%%%%%%%%%%%%%%%%%%%%
\section{Conventions}\label{app:Convention}
%%%%%%%%%%%%%%%%%%%%%%%%%%%%%%%%%%%%%%%%%%%
%%%%%%%%%%%%%%%%%%%%%%%%%%%%%%%%%%%%%%%%%%%

The 3d $\gamma$-matrices are chosen to be
\be
  \gamma_1 = \sigma_3\, ,\quad \gamma_2 = - \sigma_1\, ,\quad \gamma_3 = - \sigma_2\, ,
\ee
where $\sigma_i$ are the standard Pauli matrices. They satisfy
  \begin{equation}
    [\gamma_m,\, \gamma_n] = 2i \varepsilon_{mnp} \gamma^p\, .
  \end{equation}

  In this paper, we use commuting spinors. The product of two spinors are defined as
  \begin{equation}
    \psi \chi = \psi^\alpha C_{\alpha\beta} \chi^\beta\, ,\quad \psi \gamma_\mu \chi = \psi^\alpha (C \gamma_\mu)_{\alpha\beta} \chi^\beta\, ,
  \end{equation}
  where the indices can be raised and lowered using
  \begin{displaymath}
    C = \left(\begin{array}{cc}
                                        0 & 1\\
                                        -1 & 0
                                      \end{array} \right)
  \end{displaymath}
  is the charge conjugation matrix. The spinor bilinears of commuting spinors satisfy
  \begin{equation}
    \psi \chi = -\chi \psi\, ,\quad \psi \gamma_\mu \chi = \chi \gamma_\mu \psi\, .
  \end{equation}

%%%%%%%%%%%%%%%%%%%%%%%%%%%%%%%%%%%%%%%%%%%
%%%%%%%%%%%%%%%%%%%%%%%%%%%%%%%%%%%%%%%%%%%
\section{Special Functions}\label{app:SpecialFct}
%%%%%%%%%%%%%%%%%%%%%%%%%%%%%%%%%%%%%%%%%%%
%%%%%%%%%%%%%%%%%%%%%%%%%%%%%%%%%%%%%%%%%%%

\subsection{Multiple Gamma and Multiple Zeta Functions}

The double zeta function $\zeta_2 (s;\, x | \varepsilon_1, \varepsilon_2)$ can be viewed as the regularization of the infinite sum:
\be
  \zeta_2 (s;\, x | \varepsilon_1, \varepsilon_2) = \sum_{m, n \geq 0} (x + m \varepsilon_1 + n \varepsilon_2)^{-s}\, .
\ee
The double gamma function is defined as
\be
  \Gamma_2 (x | \varepsilon_1, \varepsilon_2) = \textrm{exp} \frac{d}{ds} \bigg|_0 \zeta_2 (s;\, x | \varepsilon_1, \varepsilon_2)\, .
\ee
It can be viewed as a regularized infinite product depending on the signs of $\varepsilon_1$, $\varepsilon_2$:
\be
  \Gamma_2 (x | \varepsilon_1, \varepsilon_2) \propto \left\{
  \begin{aligned}
    & \prod_{m, n \geq 0} (x + m \varepsilon_1 + n \varepsilon_2)^{-1}\, , & \textrm{for } \varepsilon_1 > 0, \varepsilon_2 > 0\, ; \\
    & \prod_{m, n \geq 0} \left(x + m \varepsilon_1 - (n+1) \varepsilon_2 \right)\, , & \textrm{for } \varepsilon_1 > 0, \varepsilon_2 < 0\, ; \\
    & \prod_{m, n \geq 0} \left(x - (m+1) \varepsilon_1 + n \varepsilon_2 \right)\, , & \textrm{for } \varepsilon_1 < 0, \varepsilon_2 > 0\, ; \\
    & \prod_{m, n \geq 0} \left(x - (m+1) \varepsilon_1 - (n+1) \varepsilon_2 \right)^{-1}\, , & \textrm{for } \varepsilon_1 < 0, \varepsilon_2 < 0\, .
  \end{aligned} \right.
\ee

The Barnes' multiple gamma functions $\Gamma_N (z | a_1, a_2)$ satisfy the following identity for $N \in \mathbb{N}$:
\be
  \Gamma_{N+1} (z | 1, \vec{\beta})\cdot \Gamma_{N+1} (1-z | 1, - \vec{\beta}) = e^{- \pi i \zeta_{N+1} (0, z | 1, \vec{\beta})}\, \prod_{\vec{k} \in \mathbb{N}^N} \left(1 - e^{2 \pi i (z + \vec{k}\cdot \vec{\beta})} \right)^{-1}\, ,
\ee
where $\zeta_{N+1} (0, z | 1, \vec{\beta})$ are the Barnes' multiple zeta functions. In particular, we need the special case $N=1$ in this paper:
\be\label{eq:Gamma2Id}
  \Gamma_2 (z | 1, \beta)\cdot \Gamma_2 (1-z | 1, -\beta) = e^{- \pi i \zeta_2 (0, z | 1, \beta)}\, \prod_{k \in \mathbb{N}} \left(1 - e^{2 \pi i (z + k \beta)} \right)^{-1}\, ,
\ee
with $\zeta_2 (0, z | 1, \beta)$ given explicitly by
\be
  \zeta_2 (0, z | 1, \beta) = \frac{z^2}{2 \beta} - \frac{(1+\beta) z}{2 \beta} + \frac{1 + 3 \beta + \beta^2}{12\, \beta}\, .
\ee
More details of $\Gamma_N (z | a_1, a_2)$ and $\zeta_N (z | a_1, a_2)$ can be found in \cite{friedman2004shintani}.

\subsection{Polylogarithmic Functions}

In the main text we have used some properties of the polylogarithmic functions $\textrm{Li}_2 (x)$ and $\textrm{Li}_3 (x)$. The general polylogarithmic function $\textrm{Li}_n (x)$ is defined as
\be
  \textrm{Li}_k (x) \equiv \sum_{n=1}^\infty \frac{x^n}{n^k}\, .
\ee
The functions $\textrm{Li}_2 (e^{i u})$ and $\textrm{Li}_3 (e^{i u})$ have the following properties for with $0 < \textrm{Re} (u) < 2 \pi$:
\begin{align}
  \textrm{Li}_2 (e^{i u}) + \textrm{Li}_2 (e^{- i u}) & = \frac{u^2}{2} - \pi u + \frac{\pi^2}{3}\, ,\\
  \textrm{Li}_3 (e^{i u}) - \textrm{Li}_3 (e^{- i u}) & = \frac{i}{6} u^3 - \frac{i \pi}{2} u^2 + \frac{i \pi^2}{3} u\, .
\end{align}

\subsection{$q$-Pochhammer Symbol}

The $q$-Pochhammer symbol $(a;\, q)_m$ is defined as
\be
  (a;\, q)_m \equiv \prod_{k=0}^{m-1} (1 - a\, q^k)\, .
\ee
For $q \approx 1$, the $q$-Pochhammer symbol $(a;\, q)_\infty$ has an expansion \cite{Zagier}:
\be
  \textrm{log} (z,\, e^{\hbar})_\infty = \frac{1}{\hbar} \sum\limits_{n=0}^\infty \frac{B_n\, \hbar^n}{n!} {\rm Li}_{2-n}(z) , \qquad {\rm Re}\,\hbar<0,
 \ee
where $B_n$ are the Bernoulli numbers.

\bibliographystyle{utphys}
\bibliography{ABJMmatrix}

\end{document}